\documentclass[12pt,journal,letterpaper,compsoc]{IEEEtran}
\pdfoutput=1
\IEEEoverridecommandlockouts

\usepackage{amsfonts}
\usepackage{amsmath}
\usepackage{amssymb}
\usepackage{amsthm}
\usepackage{graphicx}
\usepackage{subfigure}
\usepackage[linesnumbered,boxed,ruled,vlined]{algorithm2e}

\onecolumn
\renewcommand{\baselinestretch}{1.2}

\ifCLASSINFOpdf

\else

\fi

\begin{document}

\title{Design of a Sliding Window over Asynchronous Event Streams}

\author{\IEEEauthorblockN{Yiling Yang$^{1,2}$, Yu Huang$^{1,2}$\IEEEauthorrefmark{1}, Jiannong Cao$^3$, Xiaoxing Ma$^{1,2}$, Jian Lu$^{1,2}$\\}
\thanks{\IEEEauthorrefmark{1}Corresponding author.}
\IEEEauthorblockA{$^1$State Key Laboratory for Novel Software Technology\\
Nanjing University, Nanjing 210093, China\\
$^2$Department of Computer Science and Technology\\
Nanjing University, Nanjing 210093, China\\
csylyang@smail.nju.edu.cn,
\{yuhuang, xxm, lj\}@nju.edu.cn\\}
$^3$Internet and Mobile Computing Lab, Department of Computing\\
Hong Kong Polytechnic University, Hong Kong, China\\
csjcao@comp.polyu.edu.hk
}

\maketitle

\begin{abstract}

The proliferation of sensing and monitoring applications motivates adoption of the event stream model of computation. Though sliding windows are widely used to facilitate effective event stream processing, it is greatly challenged when the event sources are distributed and asynchronous. To address this challenge, we first show that the snapshots of the asynchronous event streams within the sliding window form a convex distributive lattice (denoted by {\it Lat-Win}). Then we propose an algorithm to maintain {\it Lat-Win} at runtime. The {\it Lat-Win} maintenance algorithm is implemented and evaluated on the open-source context-aware middleware we developed. The evaluation results first show the necessity of adopting sliding windows over asynchronous event streams. Then they show the performance of detecting specified predicates within {\it Lat-Win}, even when faced with dynamic changes in the computing environment.

\end{abstract}

\IEEEpeerreviewmaketitle

\section{Introduction}
\label{sec:introduction}

Sensing devices such as wireless sensor motes and RFID readers are gaining adoption on an increasing scale for tracking and monitoring purposes. An emerging class of applications includes context-aware computing in a smart home/office \cite{Huang09, Huang11}, supply chain management \cite{Niederman07}, and facility management \cite{Wu06}. These applications require the online processing of a large amount of events from multiple event sources, which necessitate the {\it event stream model} of computation \cite{Tirthapura06, Wu06}.

In tracking and monitoring applications, event streams are often generated from multiple distributed sources. More importantly, the event sources may not have a global clock or shared memory. Communications among the event sources may suffer from finite but arbitrary delay. It is a critical challenge how to process such {\it asynchronous event streams}  \cite{Tirthapura06, Huang09, Huang11}.

For example in a smart office scenario, the context-aware middleware may receive the event stream of user's location updates from his mobile phone (we assume that the user's location can be decided by the access point his phone connects to) \cite{Huang09, Huang11}. The middleware may also receive event streams from sensors in the meeting room about whether there is a presentation going on. Due to the asynchrony among the event sources, the middleware cannot easily decide the composite global event ``the user is in the meeting room, where a presentation is going on", in order to mute the mobile phone intelligently.

Coping with the asynchrony has been widely studied in distributed computing \cite{Babaoglu93, Schwarz94}. One important approach relies on the ``happen-before" relation resulting from message passing \cite{Lamport78}. Based on this relation, various types of logical clocks can be devised \cite{Schwarz94, Mattern89}. Based on logical time, one key notion in an asynchronous system is that all meaningful observations or global snapshots of the system form a distributive lattice \cite{Babaoglu93, Schwarz94}.

In tracking and monitoring applications, the events may quickly accumulate to a huge volume, and so will the lattice of global snapshots of the asynchronous event streams\cite{Schwarz94, Jard94}. Processing of the entire event streams is often infeasible and, more importantly, not necessary \cite{Braverman11}. In such applications, we are often concerned only on the most recent events. This can be effectively captured by the notion of a {\it sliding window} \cite{Datar02, Tirthapura06, Braverman11}. Processing events within the window (discarding the stale events) can greatly reduce the processing cost. Also in the smart office scenario, user's location half an hour ago is often of little help in meeting his current need. Thus we can keep a sliding window (say latest 5 location updates) over the user's location stream.

Challenge of the asynchrony and effectiveness of the sliding window motivate us to study the following problem. In a system of $n$ asynchronous event streams and one sliding window on each stream, we define an {\it n-dimensional sliding window} as the Cartesian product of the window on every event source. Considering the system of asynchronous event streams within the $n$-dimensional sliding window, does the lattice structure of global snapshots preserve? If it does, how to effectively maintain this lattice of snapshots at runtime? How to support effective detection of predicates over event streams within the window? Toward these problems, the contribution of this work is two-fold:
\begin{itemize}
  \item We first prove that global snapshots of asynchronous event streams within the $n$-dimensional sliding window form a distributive lattice (denoted by {\it Lat-Win}). We also find that {\it Lat-Win} is a convex sub-lattice of the ``original lattice'' (obtained when no sliding window is imposed and the entire streams are processed);
  \item Then we characterize how {\it Lat-Win} evolves when the window slides over the asynchronous event streams. Based on the theoretical characterization, we propose an online algorithm to maintain {\it Lat-Win} at runtime.
\end{itemize}

A case study of a smart office scenario is conducted to demonstrate how our proposed {\it Lat-Win} facilitates context-awareness in asynchronous pervasive computing scenarios \cite{Huang09, Huang11}. The {\it Lat-Win} maintenance algorithm is implemented and evaluated over MIPA -- the open-source context-aware middleware we developed \cite{MIPA, Huang10b, Huang11}. The performance measurements first show the necessity of adopting the sliding window over asynchronous event streams. Then the measurements show that using the sliding window, fairly accurate predicate detection (accuracy up to 95\%) can be achieved, while the cost of event processing can be greatly reduced (to less than 1\%).

The rest of this paper is organized as follows. Section \ref{sec:preliminaries} presents the preliminaries. Section \ref{sec:design-overview} overviews how {\it Lat-Win} works, while Section \ref{sec:characterization} and \ref{sec:algorithm} detail the theoretical characterization and algorithm design respectively. Section \ref{sec:evaluation} presents the experimental evaluation. Section \ref{sec:related-work} reviews the related work. Finally, In Section \ref{sec:conclusion}, we conclude the paper and discuss the future work.

\section{Preliminaries}
\label{sec:preliminaries}

In this section, we first describe the system model of asynchronous event streams. Then we discuss the lattice of global snapshots of asynchronous event streams. Finally, we introduce the {\it $n$-dimensional sliding window} over asynchronous event streams. Notations used through out this work are listed in Table \ref{T:Notations-Model}.
\begin{table}[htbp]
\caption{Notations Used in Design of {\it Lat-Win}}
\label{T:Notations-Model} \centering
\begin{tabular}{r | l}
\hline
Notation & Explanation \\
\hline
\hline

$n$ & number of non-checker processes \\

$P^{(k)}, P_{che}$ & non-checker / checker process ($1\leq k\leq n$) \\

$e^{(k)}_i, s^{(k)}_i$ & event / local state on $P^{(k)}$ \\

$Que^{(k)}$ & queue of local states from each $P^{(k)}$ on $P_{che}$ \\

$W^{(k)}$ & sliding window on a single event stream \\

$W^{(k)}_{min} / W^{(k)}_{max}$ & the oldest/latest local state within $W^{(k)}$ \\

$w$ & uniform size of every $W^{(k)}$ \\

$W$ & $n$-dimensional sliding window over asynchronous event streams \\

$\mathcal{G}$ & global state of the asynchronous event streams \\

$\mathcal{C}$ & Consistent Global State (CGS) \\

$\mathcal{C}[k]$ & $k^{th}$ constituent local state of $\mathcal{C}$ \\

{\it LAT} & original lattice of CGSs when no sliding window is used and the entire streams are processed \\

{\it Lat-Win} & lattice of CGSs within the $n$-dimensional sliding window \\

$\mathcal{C}_{min}, \mathcal{C}_{max}$ &  the minimal/maximal CGSs in {\it Lat-Win} \\

\hline
\end{tabular}
\end{table}

\subsection{A System of Asynchronous Event Streams}

\begin{figure}[htbp]
    \subfigure[Sliding windows over asynchronous event streams]{
        \label{F:space time diagram}
        \begin{minipage}[b]{0.5\textwidth}
        \centering
        \includegraphics[width=2.5in]{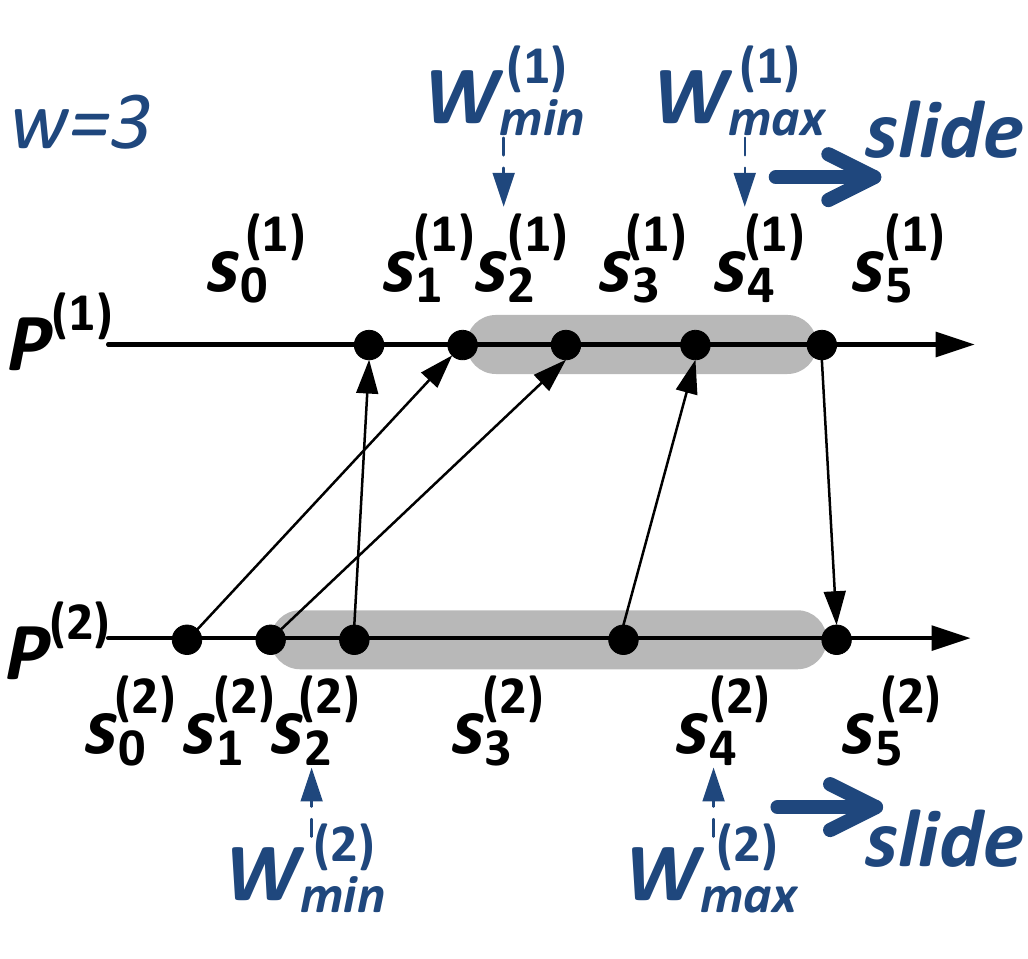}
        \end{minipage}}\hspace*{\fill}
    \subfigure[The $n$-dimensional sliding window over the lattice]{
        \label{F:window over lattice}
        \begin{minipage}[b]{0.5\textwidth}
        \centering
        \includegraphics[width=2.5in]{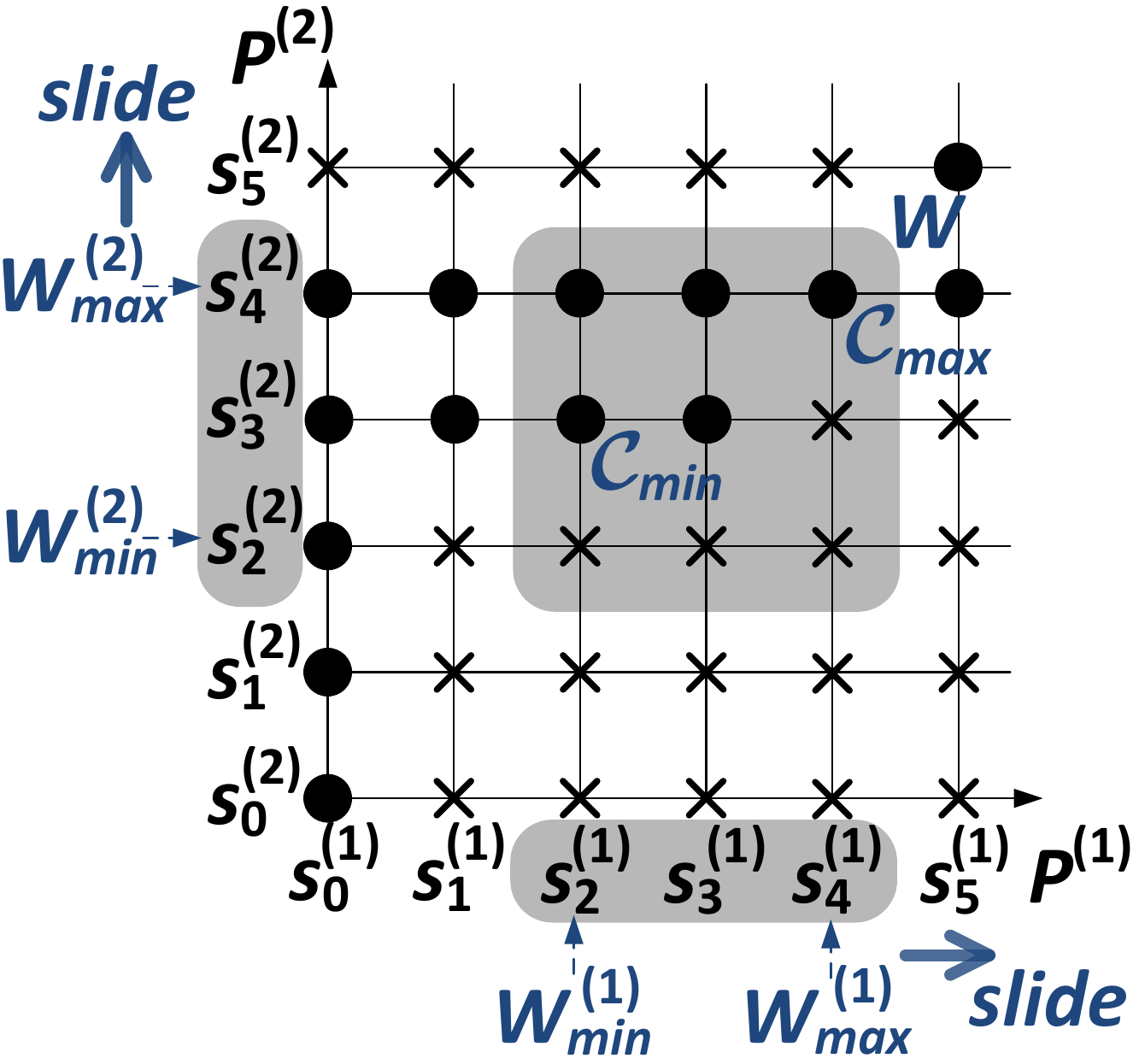}
        \end{minipage}}\\
    \centering\parbox[c]{1.6in}{\caption{System model}}
    \label{F:system model}
\end{figure}

In a tracking/monitoring application, we are faced with multiple distributed event sources which generate event streams at runtime. The event sources do not necessarily have global clocks or shared memory. The event sources are modeled as $n$ {\it non-checker processes} $P^{(1)}, P^{(2)}, \cdots, P^{(n)}$. Each $P^{(k)}$ produces a stream of {\it events} connected by its {\it local states}: ``$e^{(k)}_0$, $s^{(k)}_0$, $e^{(k)}_1$, $s^{(k)}_1$, $e^{(k)}_2$, $\cdots$", as shown in Fig. \ref{F:space time diagram}. The event may be local, indicating status update of the entity being monitored and causing a local state change, or global, e.g. communication via sending/receiving messages. The non-checker processes form a loosely-coupled asynchronous system. We assume that no messages are lost, altered or spuriously introduced, as in \cite{Garg94, Garg96}. The underlying communication channel is not necessarily FIFO.

We re-interpret the notion of time based on Lamport's definition of the {\it happen-before} relation (denoted by `$\rightarrow$') resulting from message causality \cite{Lamport78}. This happen-before relation can be effectively encoded and decoded based on the logical vector clock scheme \cite{Mattern89}. Specifically, for two events $e^{(i)}_a$ and $e^{(j)}_b$ in the system of asynchronous event streams, we have $e^{(i)}_a \rightarrow e^{(j)}_b$ iff:

\begin{itemize}
  \item $(i = j) \wedge (b = a + 1)$, or
  \item $(e^{(i)}_a = send(m)) \wedge (e^{(j)}_b = receive(m))$, or
  \item $\exists \ e^{(k)}_c : (e^{(i)}_a\rightarrow e^{(k)}_c) \wedge (e^{(k)}_c\rightarrow e^{(j)}_b)$.
\end{itemize}

\noindent For two local states $s_1$ and $s_2$, $s_1 \rightarrow s_2$ iff the ending of $s_1$ happen-before (or coincides with) the beginning of $s_2$ (note that the beginning and ending of a state are both events). As shown in Fig. \ref{F:space time diagram}, $s^{(2)}_{2} \rightarrow s^{(1)}_{1}$ and $s^{(1)}_{4} \rightarrow s^{(2)}_{5}$.

One {\it checker process} $P_{che}$ is in charge of collecting and processing the asynchronous event streams. For example in a context-aware computing scenario \cite{Huang11}, $P_{che}$ may be a context reasoning process deployed over the context-aware middleware. In a supply chain management scenario, $P_{che}$ may be a central administration application, monitoring the progresses of multiple supply chains.

Whenever $P^{(k)}$ generates a new event and proceeds to a new local state, it sends the local state with the vector clock timestamp to $P_{che}$. We use message sequence numbers to ensure that $P_{che}$ receives messages from each $P^{(k)}$ in FIFO manner \cite{Garg94, Garg96, Huang09, Huang11}.

\subsection{Lattice of Consistent Global States(CGS)}
\label{sec:lattice}

In the tracking/monitoring application, we are concerned with the state of the entities being monitored after specific events are executed. For a system of asynchronous event streams, we are thus concerned with the global states or snapshots of the whole system.

A global state $\mathcal{G} = (s^{(1)}, s^{(2)}, \cdots, s^{(n)})$ of asynchronous event streams is defined as a vector of local states from each non-checker process $P^{(k)}$. A global state may be either consistent or inconsistent. The notion of {\it Consistent Global State (CGS)} is crucial in processing of asynchronous event streams. Intuitively, a global state is consistent if an omniscient external observer could actually observe that the system enters that state. Formally, a global state $\mathcal{C}$ is {\it consistent} if and only if the constituent local states are pairwise concurrent \cite{Babaoglu93}, i.e., $$\mathcal{C} = (s^{(1)}, s^{(2)}, \cdots, s^{(n)}), \forall\ i,j : i\neq j :: \neg (s^{(i)}\rightarrow s^{(j)})$$

\noindent The CGS denotes a global snapshot or meaningful observation of the system of asynchronous event streams.

It is intuitive to define the {\it precede} relation (denoted by `$\prec$') between two CGSs: $\mathcal{C} \prec \mathcal{C'}$ if $\mathcal{C'}$ is obtained via advancing $\mathcal{C}$ by exactly one local state on one non-checker process. The {\it lead-to} relation (denoted by `$\leadsto$') is defined as the transitive closure of `$\prec$'.

The set of all CGSs together with the `$\leadsto$' relation form a distributive lattice \cite{Babaoglu93, Schwarz94}. As shown in Fig. \ref{F:window over lattice}, black dots denote the CGSs and the edges between them depict the '$\prec$' relation. The crosses ``$\times$'' denote the inconsistent global states. The lattice structure among all CGSs serves as a key notion for the detection of global predicates over asynchronous event streams \cite{Babaoglu93, Schwarz94}.

\subsection{The $n$-dimensional Sliding Window over Asynchronous Event Streams}

On $P_{che}$, states of each event source $P^{(k)}$ are queued in $Que^{(k)}$. As discussed in Section \ref{sec:introduction}, in many cases, it is too expensive and often unnecessary to process the entire event stream. A {\it local sliding window} $W^{(k)}$ of size $w$ is imposed on each $Que^{(k)}$. Then we can define the {\it $n$-dimensional sliding window} $W$ as the Cartesian product of each $W^{(k)}$: $W = W^{(1)} \times W^{(2)} \times \cdots \times W^{(n)}$.

As shown in Fig. \ref{F:space time diagram}, the window $W^{(1)}$ with $w = 3$ on $P^{(1)}$ currently contains \{$s^{(1)}_{2}$, $s^{(1)}_{3}$, $s^{(1)}_{4}$\}. The 2-dimensional sliding window $W^{(1)}\times W^{(2)}$ is depicted by the gray square in Fig. \ref{F:window over lattice}. The arrival of $s^{(1)}_{5}$ will trigger the 2-dimensional window to slide in $P^{(1)}$'s dimension, and $W^{(1)}$ is updated to \{$s^{(1)}_{3}$, $s^{(1)}_{4}$, $s^{(1)}_{5}$\}.

We assume that the concurrency control scheme is available on $P_{che}$, which means that the events from all non-checker processes are processed one at a time. We also assume that the sliding windows on the event streams have uniform size $w$. Note that this assumption is not restrictive and is for the ease of interpretation. Our proposed scheme also works if the windows on different streams have different sizes.

\section{Lat-Win - Design Overview}
\label{sec:design-overview}

The central problem in this work is how to characterize and maintain the $n$-dimensional sliding window over asynchronous event streams. Toward this problem, our contribution is two-fold. First, we characterize {\it Lat-Win} - the lattice of CGSs over the asynchronous event streams within the $n$-dimensional sliding window. Then we propose an online algorithm to maintain {\it Lat-Win} at runtime.
\begin{figure}[htbp]
        \begin{minipage}[b]{1\textwidth}
        \centering
        \includegraphics[width=7in]{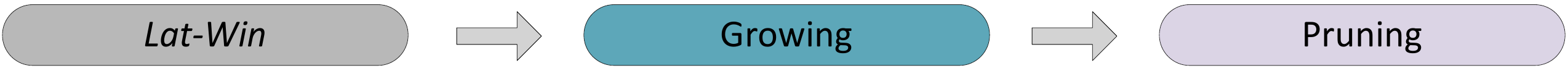}
        \end{minipage}\\
    \subfigure[{\it Lat-Win} induced by \{$s^{(1)}_2$, $s^{(1)}_3$, $s^{(1)}_4$, $s^{(2)}_2$, $s^{(2)}_3$, $s^{(2)}_4$\}, and $s^{(1)}_5$ arrives]{
        \label{F:design overview 1}
        \begin{minipage}[b]{0.32\textwidth}
        \centering
        \includegraphics[width=2.2in]{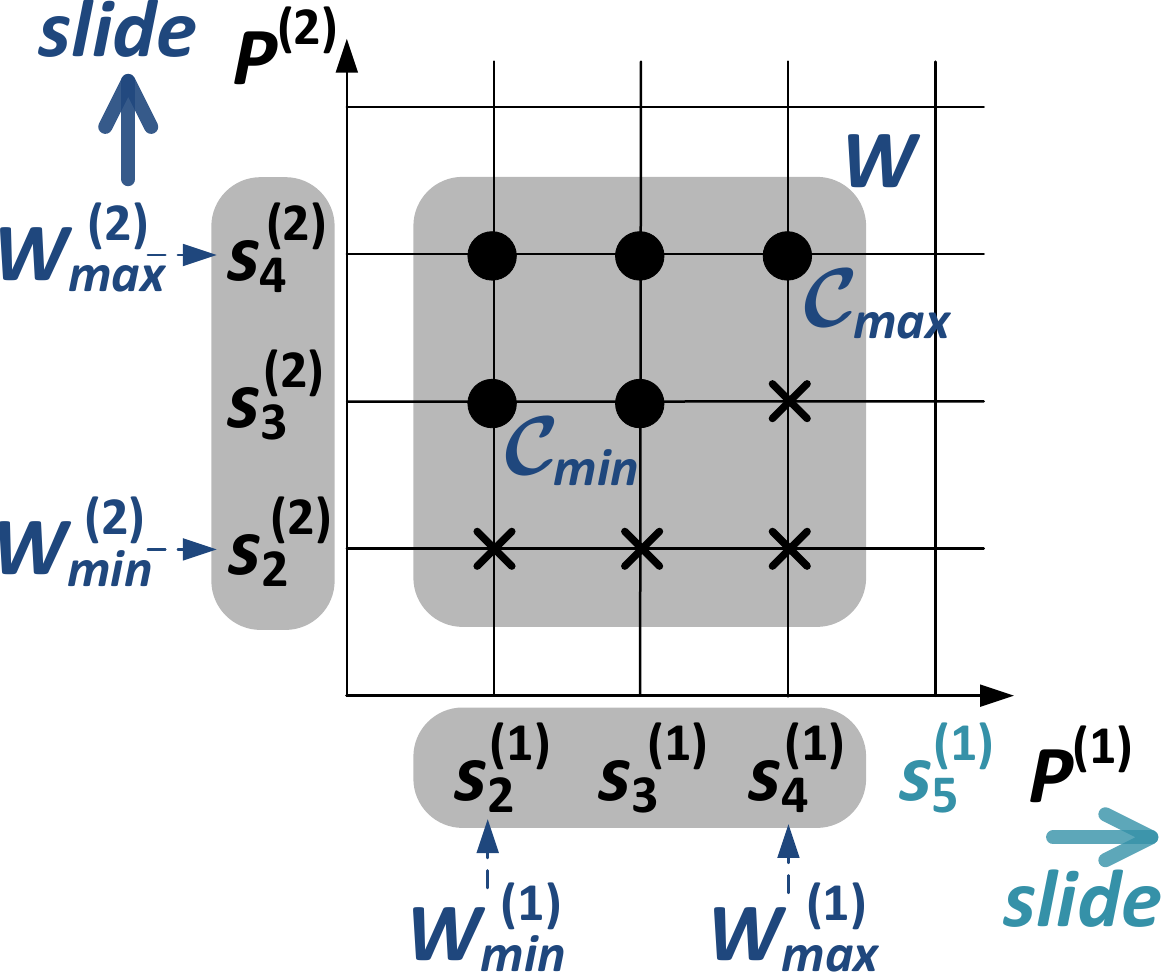}
        \end{minipage}}\hspace*{\fill}
    \subfigure[{\it Lat-Win} grows with the CGSs which contain $s^{(1)}_5$ as a constituent, and $\mathcal{C}_{max}$ is updated]{
        \label{F:design overview 2}
        \begin{minipage}[b]{0.32\textwidth}
        \centering
        \includegraphics[width=2.2in]{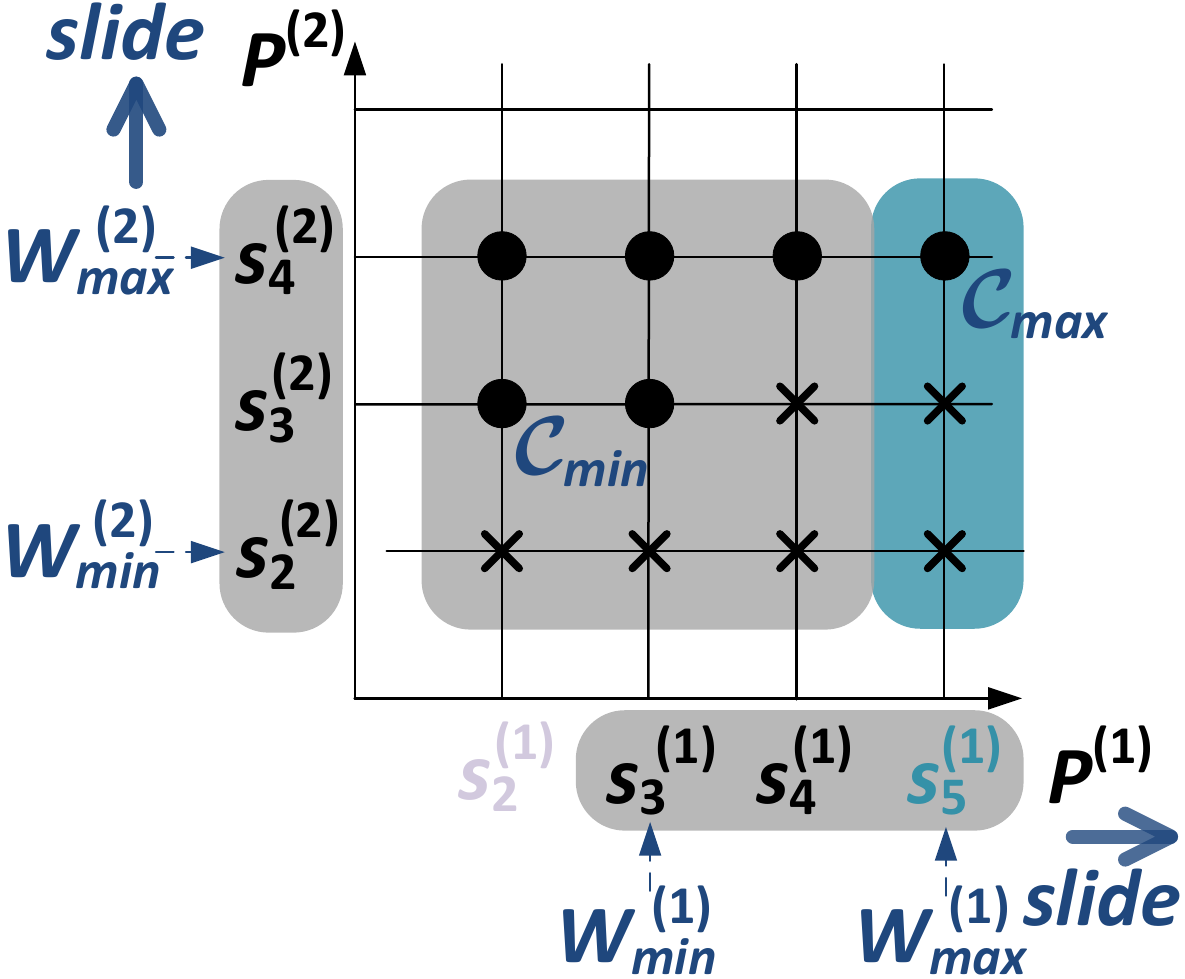}
        \end{minipage}}\hspace*{\fill}
    \subfigure[{\it Lat-Win} prunes the CGSs which contain $s^{(1)}_2$ as a constituent, and $\mathcal{C}_{min}$ is updated]{
        \label{F:design overview 3}
        \begin{minipage}[b]{0.32\textwidth}
        \centering
        \includegraphics[width=2.2in]{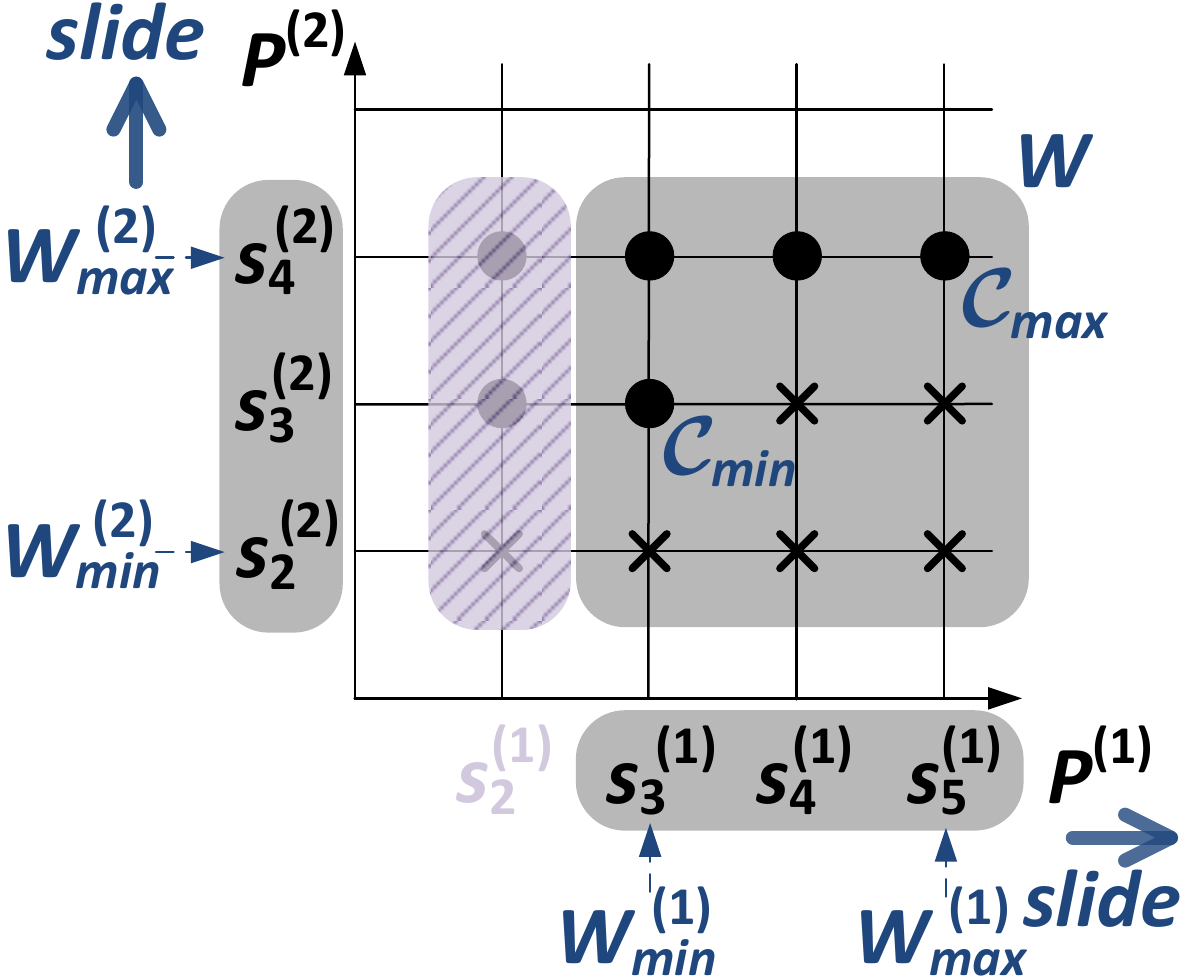}
        \end{minipage}}
    \caption{Online maintenance of {\it Lat-Win}. When $s^{(1)}_5$ arrives, {\it Lat-Win} first grows with a set of new CGSs and then prunes the stale CGSs.}
    \label{F:design overview}
\end{figure}

\subsection{Characterization of Lat-Win}

An important property concerning {\it Lat-Win} is that all CGSs within the $n$-dimensional sliding window together with the '$\leadsto$' relation have the lattice structure. Moreover, {\it Lat-Win} turns out to be a distributive convex sublattice of the original lattice {\it LAT} (the lattice obtained when no sliding window is used and all event streams are processed). As shown in Fig. \ref{F:window over lattice}, the gray square in the middle is a 2-dimensional sliding window over two asynchronous event streams produced by $P^{(1)}$ and $P^{(2)}$. The CGSs within the square form a convex sublattice of the original lattice, i.e., the {\it Lat-Win}.

When an event $e^{(i)}_{j}$ is executed and $P^{(i)}$ arrives at a new local state $s^{(i)}_j$, the stale local state $s^{(i)}_k$ ($j - k = w$) in window $W^{(i)}$ will be discarded. The {\it Lat-Win} will ``grow'' with a set of CGSs consisting of $s^{(i)}_j$ and other local states from $W^{(m)} (m \neq i)$, and ``prune'' the CGSs which contain $s^{(i)}_k$ as a constituent.

For example in Fig. \ref{F:design overview}, assume that the {\it Lat-Win} is initially shown in Fig. \ref{F:design overview 1}. When a new local state $s^{(1)}_{5}$ arrives, $s^{(1)}_2$ will be discarded. State $s^{(1)}_{5}$ will be combined with local states in $W^{(2)}$ to obtain the CGS $\mathcal{C}_{5,4} = (s^{(1)}_5, s^{(2)}_4)$ in the blue rectangle in Fig. \ref{F:design overview 2}. CGSs which contain $s^{(1)}_2$ as a constituent in the left shaded rectangle will be discarded as shown in Fig. \ref{F:design overview 3}. The CGSs in the current window (e.g., the gray square in Fig. \ref{F:design overview 3}) remain to be a sublattice. It seems that the 2-dimensional window containing the {\it Lat-Win} slides over the asynchronous event streams produced by $P^{(1)}$ and $P^{(2)}$.

\subsection{Online Maintenance of Lat-Win}

Based on the theoretical characterization above, we propose an algorithm for maintaining {\it Lat-Win} at runtime. Let $\mathcal{C}_{min}$ ($\mathcal{C}_{max}$) denote the CGS which has no predecessors (successors) in {\it Lat-Win}. $\mathcal{C}_{min}$ and $\mathcal{C}_{max}$ serve as two ``anchors" in updating {\it Lat-Win}. When a new local state arrives, the {\it Lat-Win} ``grows" from $\mathcal{C}_{max}$ and ``prunes" from $\mathcal{C}_{min}$, as shown in Fig. \ref{F:design overview}. After the growing and pruning, $\mathcal{C}_{min}$ and $\mathcal{C}_{max}$ are also updated for further maintenance of {\it Lat-Win}. Due to the symmetry in the lattice structure, the growing and pruning of {\it Lat-Win} are dual. So are the updates of $\mathcal{C}_{min}$ and $\mathcal{C}_{max}$.

\section{Lat-Win - Characterizing the Snapshots of Windowed Asynchronous Event Streams}
\label{sec:characterization}

The theoretical characterization of {\it Lat-Win} consists of two parts. First we study the lattice of snapshots within the $n$-dimensional sliding window. Then we study how the {\it Lat-Win} evolves as the $n$-dimensional window slides.

\subsection{Sub-lattice within the Sliding Window}

An $n$-dimensional sliding window consists of $n$ local windows sliding on event streams produced by non-checker processes $P^{(1)}, P^{(2)}, \cdots, P^{(n)}$, and induces $n$ segments of local states $W^{(1)}, W^{(2)}, \cdots , W^{(n)}$.

The happen-before relation between local states has been encoded in their logical clock timestamps. Based on the local states as well as the happen-before relation among them, we can get a set of CGSs within the $n$-dimensional sliding window. An important property we find is that the CGSs within the $n$-dimensional sliding window, together with the `$\leadsto$' relation, also form a lattice - {\it Lat-Win}. More importantly, {\it Lat-Win} is a {\it distributive convex sub-lattice} of the original lattice {\it LAT}. Formally,\\

\noindent{\bf Theorem 1.} {\it Given an $n$-dimensional sliding window $W = W^{(1)} \times W^{(2)} \times \cdots \times W^{(n)}$ over asynchronous event streams, let $Set_{\mathcal{C}}(W)$ denote the CGSs constructed from local states in $W$. If $Set_{\mathcal{C}}(W)$ is not empty,
\begin{itemize}
    \item[1.] ($Set_{\mathcal{C}}(W)$, $\leadsto$) forms a lattice, denoted by {\it Lat-Win};
    \item[2.] {\it Lat-Win} is a sublattice of {\it LAT};
    \item[3.] {\it Lat-Win} is convex and distributive.
\end{itemize}}

\noindent{\it Proof:}

1.1: A {\it lattice} is a poset $L$ such that for all $x, y \in L$, the least upper bound ({\it join}) of $x$ and $y$ (denoted $x \sqcup y$) and the greatest lower bound ({\it meet}) of $x$ and $y$ (denoted $x \sqcap y$) exist and are contained in the poset. For two CGSs $\mathcal{C}_i, \mathcal{C}_j, \mathcal{C}_i \sqcap \mathcal{C}_j = (min(\mathcal{C}_i[1], \mathcal{C}_j[1]), \cdots, min(\mathcal{C}_i[n], \mathcal{C}_j[n]))$, $\mathcal{C}_i \sqcup \mathcal{C}_j = (max(\mathcal{C}_i[1], \mathcal{C}_j[1]), \cdots, max(\mathcal{C}_i[n], \mathcal{C}_j[n]))$.

We prove it by contradiction. Assume that $\exists \mathcal{C}_i, \mathcal{C}_j\in Set_{\mathcal{C}}(W)$ and $\mathcal{C}_i \sqcap \mathcal{C}_j$ does not exist. It is obvious that $\mathcal{C}_i \sqcap \mathcal{C}_j$ is unique, so as $\mathcal{C}_i \sqcup \mathcal{C}_j$. It is to say that $\mathcal{C}_i \sqcap \mathcal{C}_j$ is not a CGS, that is, $\exists s,t, min(\mathcal{C}_i[s], \mathcal{C}_j[s]) \rightarrow min(\mathcal{C}_i[t], \mathcal{C}_j[t])$. Assume without loss of generality that $min(\mathcal{C}_i[s], \mathcal{C}_j[s]) = \mathcal{C}_i[s]$, i.e., $\mathcal{C}_i[s] \rightarrow \mathcal{C}_j[s]$ or $\mathcal{C}_i[s] = \mathcal{C}_j[s]$. Then, we get $\mathcal{C}_i[s] \rightarrow min(\mathcal{C}_i[t], \mathcal{C}_j[t])$. Thus, $\mathcal{C}_i[s] \rightarrow \mathcal{C}_i[t]$, which is contrary to that $\mathcal{C}_i$ is a CGS. Thus, $\mathcal{C}_i \sqcap \mathcal{C}_j$ exists. The proof of the existence of $\mathcal{C}_i \sqcup \mathcal{C}_j$ is the same.

It is easy to prove that $\mathcal{C}_i \sqcap \mathcal{C}_j$ and $\mathcal{C}_i \sqcup \mathcal{C}_j$ are both in $Set_{\mathcal{C}}(W)$, because the constituent local states of $\mathcal{C}_i \sqcap \mathcal{C}_j$ and $\mathcal{C}_i \sqcup \mathcal{C}_j$ are all in $W^{(1)}, W^{(2)}, \cdots, W^{(n)}$, and $Set_{\mathcal{C}}(W)$ contains all the CGSs constructed from local states in $W^{(1)}, W^{(2)}, \cdots, W^{(n)}$. Thus, ($Set_{\mathcal{C}}(W)$, $\leadsto$) forms a lattice.

1.2: Let $Set_{\mathcal{C}}(LAT)$ denote the CGSs of the original lattice {\it LAT}. A subset $S\subseteq L$, is a {\it sublattice} of lattice $L$, iff $S$ is non-empty and $\forall a,b\in S$, $((a\sqcap b)\in S) \wedge ((a \sqcup b)\in S)$. It is obvious that $Set_{\mathcal{C}}(W)$ of {\it Lat-Win} is a subset of $Set_{\mathcal{C}}(LAT)$. From the proof of Theorem 1.1, we can easily prove that {\it Lat-Win} is a sublattice of {\it LAT}.

1.3: A subset $S$ of a lattice $L$ is called {\it convex} iff $\forall a,b \in S, c \in L$, and $a\leq c\leq b$ imply that $c\in S$ (see Section I.3 in \cite{Gratzer03}). For three CGSs $\mathcal{C}_i, \mathcal{C}_j \in Set_{\mathcal{C}}(W)$, $\mathcal{C}_k \in Set_{\mathcal{C}}(LAT)$, $\mathcal{C}_i \leadsto \mathcal{C}_k \leadsto \mathcal{C}_j$, it infers that $\forall t, (\mathcal{C}_i[t] \rightarrow \mathcal{C}_k[t]$ or $\mathcal{C}_i[t] = \mathcal{C}_k[t]) \wedge (\mathcal{C}_k[t] \rightarrow \mathcal{C}_j[t]$ or $\mathcal{C}_k[t] = \mathcal{C}_j[t])$. Note that $\mathcal{C}_i[t], \mathcal{C}_j[t] \in W^{(t)}$ and $W^{(t)}$ contains all local states within $[\mathcal{C}_i[t], \mathcal{C}_j[t]]$. Thus, $\mathcal{C}_k[t]\in W^{(t)}$, and $\mathcal{C}_k\in Set_{\mathcal{C}}(W)$. Thus, {\it Lat-Win} is a convex sublattice of the original lattice {\it LAT}.

It is a well known result in lattice theory \cite{Davey02} that the set of all CGSs of a distributed computation forms a distributive lattice under the $\subseteq$ relation. Thus, {\it LAT} is a distributive lattice. It can be proved that any sublattice of a distributive lattice is also a distributive lattice \cite{Davey02}. Thus, {\it Lat-Win} is also a distributive lattice. \qed \\

The geometric interpretation of Theorem 1 is that $W$ can be viewed as an $n$-dimensional ``cube'' over the original lattice, and CGSs within this cube also form a lattice {\it Lat-Win}. Moreover, the `convex' and `distributive' properties of the original lattice {\it LAT} preserve when we focus on CGSs within the cube. Let $\mathcal{C}_{i,j}$ = ($s^{(1)}_{i}$, $s^{(2)}_{j}$). As shown in Fig. \ref{F:design overview 1}, the local windows are $W^{(1)}$ = \{$s^{(1)}_{2}$, $s^{(1)}_{3}$, $s^{(1)}_{4}$\} and $W^{(2)}$ = \{$s^{(2)}_{2}$, $s^{(2)}_{3}$, $s^{(2)}_{4}$\}. They define a square on the original lattice (Fig. \ref{F:window over lattice}) and induce a sublattice {\it Lat-Win} = (\{$\mathcal{C}_{2,3}$, $\mathcal{C}_{2,4}$, $\mathcal{C}_{3,3}$, $\mathcal{C}_{3,4}$, $\mathcal{C}_{4,4}$\}, $\leadsto$). The induced {\it Lat-Win} is convex because all CGSs ``greater than'' $\mathcal{C}_{2,3}$ and ``smaller than'' $\mathcal{C}_{4,4}$ in the original lattice are contained in the {\it Lat-Win}.

Given {\it Lat-Win} defined in Theorem 1, we further study how {\it Lat-Win} is contained in the cube. Is this cube a tight wrapper, i.e., does {\it Lat-Win} span to the boundary of the cube? First note that the maximal CGS and the minimal CGS are both important to the update of {\it Lat-Win}. Intuitively, the maximal CGS $\mathcal{C}_{max}$ of {\it Lat-Win} is on the upper bound $W^{(i)}_{max}$ of at least one local window $W^{(i)}$, so that {\it Lat-Win} could grow with newly arrived local states from $P^{(i)}$. Dually, the minimal CGS $\mathcal{C}_{min}$ of the {\it Lat-Win} is on the lower bound $W^{(j)}_{min}$ of at least one local window $W^{(j)}$, so that {\it Lat-Win} could grow from the stale local states from $P^{(j)}$ in the past. Formally,\\

\noindent{\bf Theorem 2.} {\it If {\it Lat-Win} is not empty,
\begin{itemize}
    \item[1.] $\exists i, \mathcal{C}_{max}[i] = W^{(i)}_{max}$;
    \item[2.] $\exists j, \mathcal{C}_{min}[j] = W^{(j)}_{min}$.
\end{itemize}}

\noindent{\it Proof:}

2.1: Let $S_{succ}(s^{(i)}_j)$ ($E_{succ}(s^{(i)}_j)$) denote the successor local state (event) to local state $s^{(i)}_j$ on $P^{(i)}$, i.e., $S_{succ}(s^{(i)}_j) = s^{(i)}_{j+1}$, $E_{succ}(s^{(i)}_j) = e^{(i)}_{j+1}$. Let $sub(\mathcal{G}, i)$ denote the global state formed by combining global state $\mathcal{G}$ and $S_{succ}(\mathcal{G}[i])$ (i.e., $sub(\mathcal{G}, i)[i] = S_{succ}(\mathcal{G}[i])$, $\forall j\neq i, sub(\mathcal{G}, i)[j] = \mathcal{G}[j]$).

We prove it by contradiction. If {\it Lat-Win} is not empty and $\forall i, \mathcal{C}_{max}[i] \neq W^{(i)}_{max}$, then $\forall i, \exists S_{succ}(\mathcal{C}_{max}[i]) \in W^{(i)}$. Because $\mathcal{C}_{max}$ is the maximal CGS, $\forall i$, global state $sub(\mathcal{C}_{max}, i)$ is not CGS.

Global state $sub(\mathcal{C}_{max}, i)$ is not CGS, $\exists j\not\in \{i\}, \mathcal{C}_{max}[j]\rightarrow S_{succ}(\mathcal{C}_{max}[i])$ and $E_{succ}(\mathcal{C}_{max}[j]) \rightarrow E_{succ}(\mathcal{C}_{max}[i])$. Global state $sub(\mathcal{C}_{max}, j)$ is not CGS, $\exists k\not\in \{i, j\}, \mathcal{C}_{max}[k]\rightarrow S_{succ}(\mathcal{C}_{max}[j])$ and $E_{succ}(\mathcal{C}_{max}[k]) \rightarrow E_{succ}(\mathcal{C}_{max}[j])$. (If $k \in \{i, j\}$, we can get that $E_{succ}(\mathcal{C}_{max}[i]) \rightarrow E_{succ}(\mathcal{C}_{max}[j]) \rightarrow E_{succ}(\mathcal{C}_{max}[i])$ or $E_{succ}(\mathcal{C}_{max}[j]) \rightarrow E_{succ}(\mathcal{C}_{max}[j])$, which is contrary to irreflexivity). By induction on the length of the set containing the used indexes (\{$i, j$\} above), we can get that to the last global state $sub(\mathcal{C}_{max}, m)$, the set contains all the indexes, and $\not\exists t\in \{1, 2, \cdots, n\}, E_{succ}(\mathcal{C}_{max}[t]) \rightarrow E_{succ}(\mathcal{C}_{max}[m])$ (If $t\in \{1, 2, \cdots, n\}$, it will lead to the contradiction to irreflexivity). Thus, if {\it Lat-Win} is not empty, $\exists i, \mathcal{C}_{max}[i] = W^{(i)}_{max}$.

2.2: The proof is dual as above. \qed \\

As shown in Fig. \ref{F:design overview 1}, the maximal CGS $\mathcal{C}_{4,4}[1]$ = $s^{(1)}_4$ = $W^{(1)}_{max}$, $\mathcal{C}_{4,4}[2]$ = $s^{(2)}_4$ = $W^{(2)}_{max}$ and the minimal CGS $\mathcal{C}_{2,3}[1]$ = $s^{(1)}_2$ = $W^{(1)}_{min}$.

\subsection{Update of Lat-Win when the Window Slides}

\begin{figure}[htbp]
    \subfigure[The restrictions which have no CGSs induced by $\mathcal{G}_{1,2}$ and $\mathcal{G}_{4,5}$]{
        \label{F:restrictions}
        \begin{minipage}[b]{0.235\textwidth}
        \centering
        \includegraphics[width=1.8in]{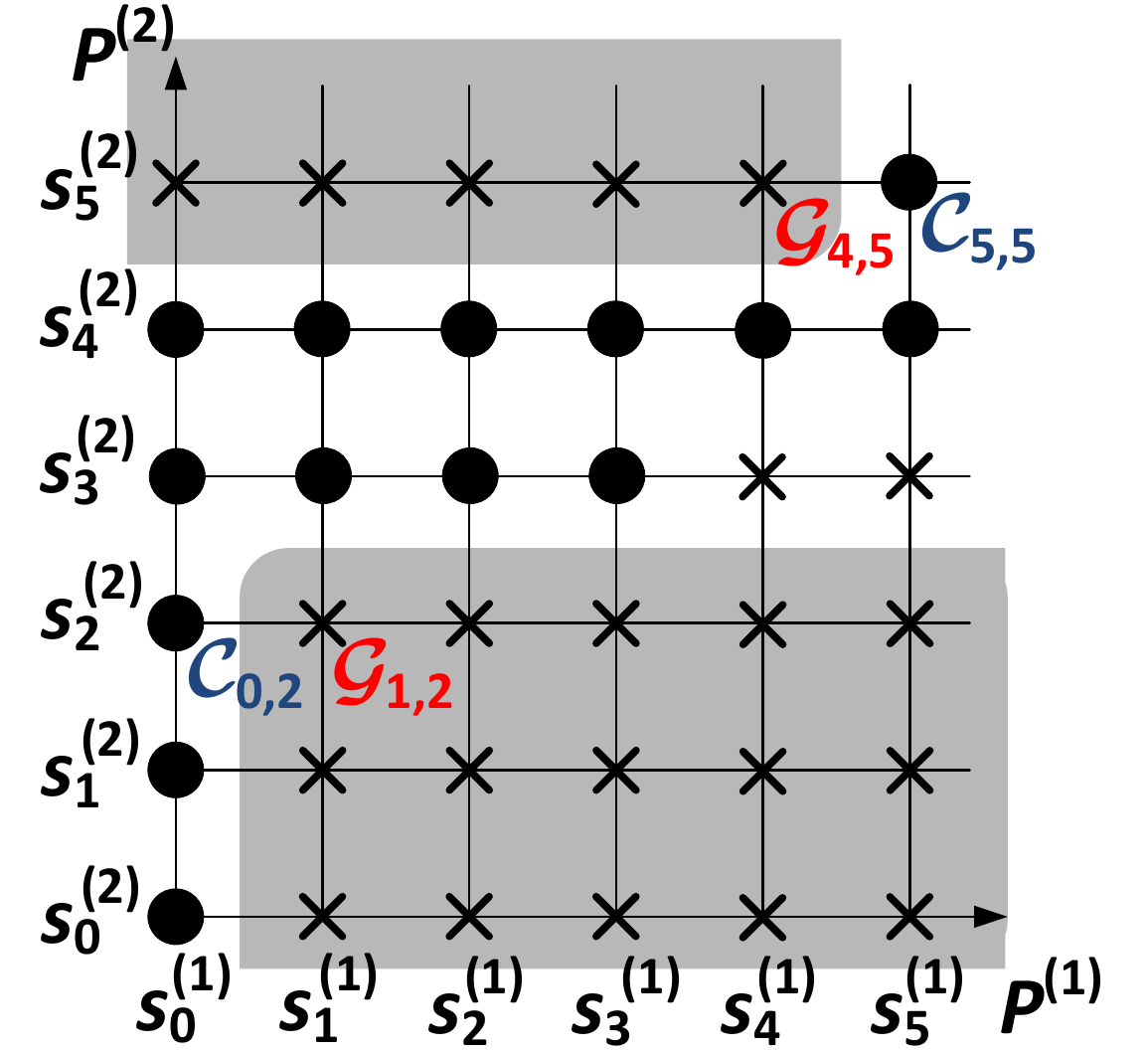}
        \end{minipage}}\hspace*{\fill}
    \subfigure[The empty window induced by \{$s^{(1)}_1$, $s^{(1)}_2$, $s^{(1)}_3$, $s^{(2)}_0$, $s^{(2)}_1$, $s^{(2)}_2$\}, and $s^{(2)}_3$ arrives]{
        \label{F:example 1}
        \begin{minipage}[b]{0.235\textwidth}
        \centering
        \includegraphics[width=1.8in]{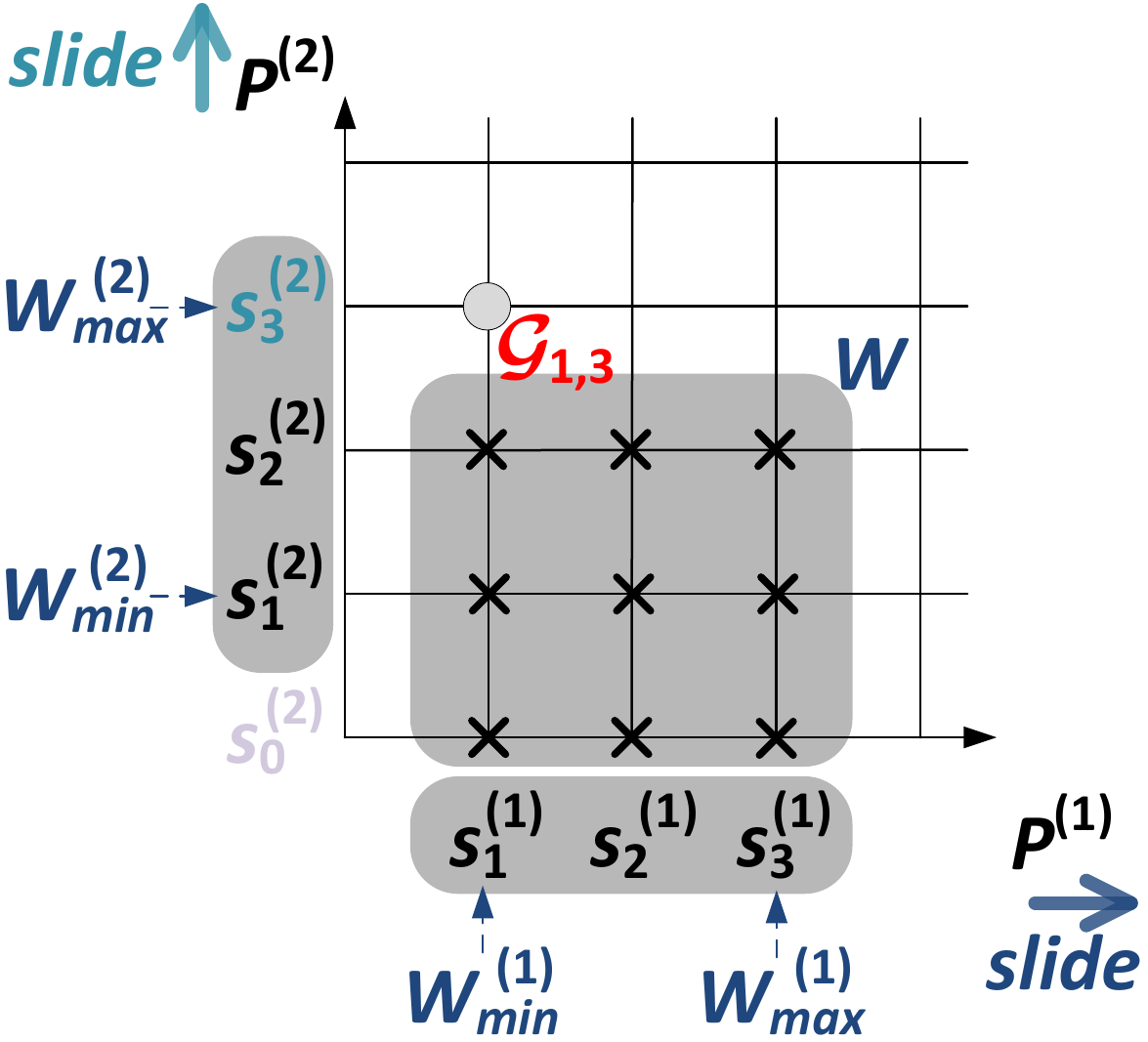}
        \end{minipage}}\hspace*{\fill}
    \subfigure[{\it Lat-Win} induced by \{$s^{(1)}_1$, $s^{(1)}_2$, $s^{(1)}_3$, $s^{(2)}_1$, $s^{(2)}_2$, $s^{(2)}_3$\}, and $s^{(1)}_4$ arrives]{
        \label{F:example 2}
        \begin{minipage}[b]{0.235\textwidth}
        \centering
        \includegraphics[width=1.8in]{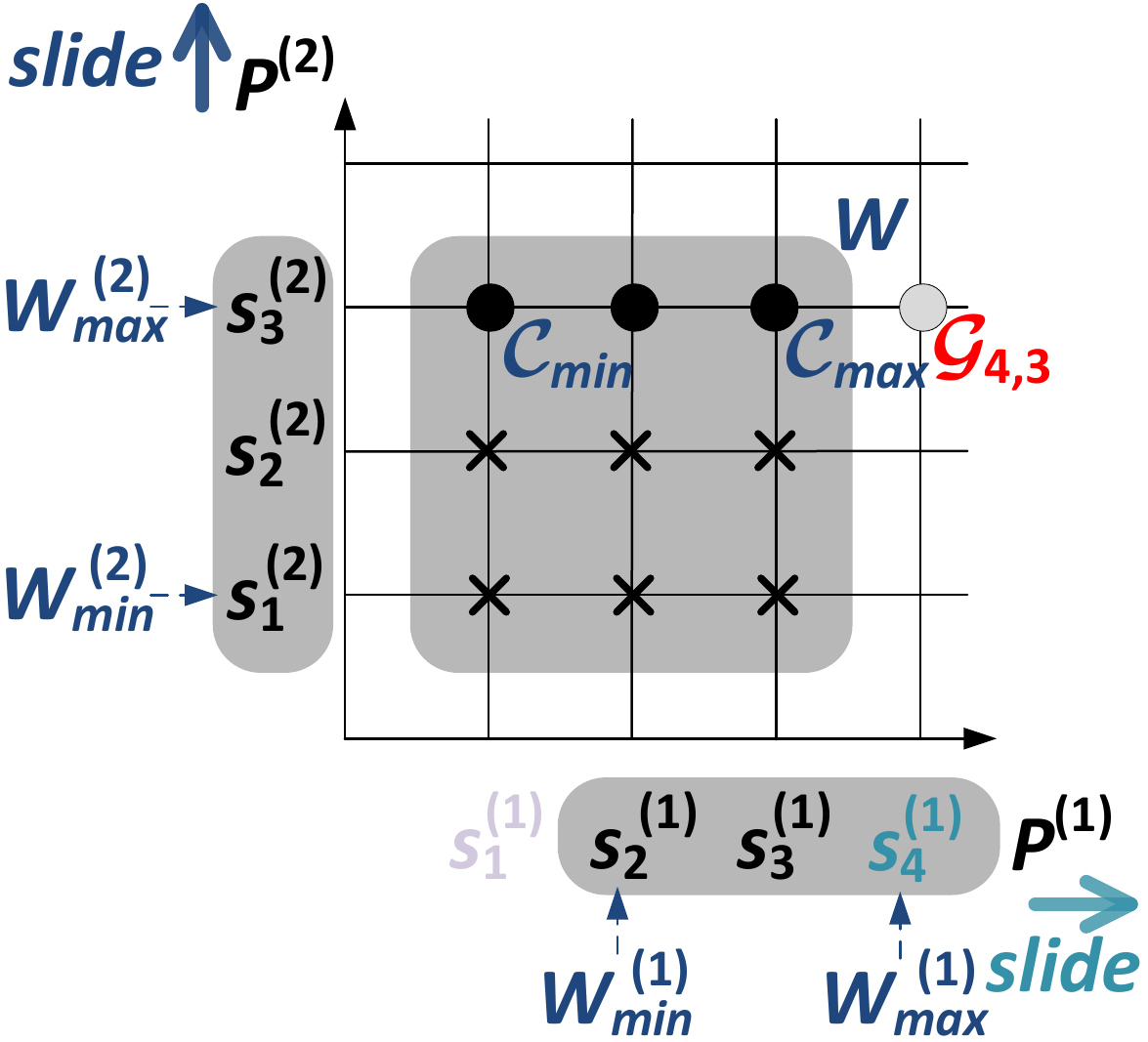}
        \end{minipage}}\hspace*{\fill}
    \subfigure[{\it Lat-Win} induced by \{$s^{(1)}_2$, $s^{(1)}_3$, $s^{(1)}_4$, $s^{(2)}_1$, $s^{(2)}_2$, $s^{(2)}_3$\}, and $s^{(2)}_4$ arrives]{
        \label{F:example 3}
        \begin{minipage}[b]{0.235\textwidth}
        \centering
        \includegraphics[width=1.8in]{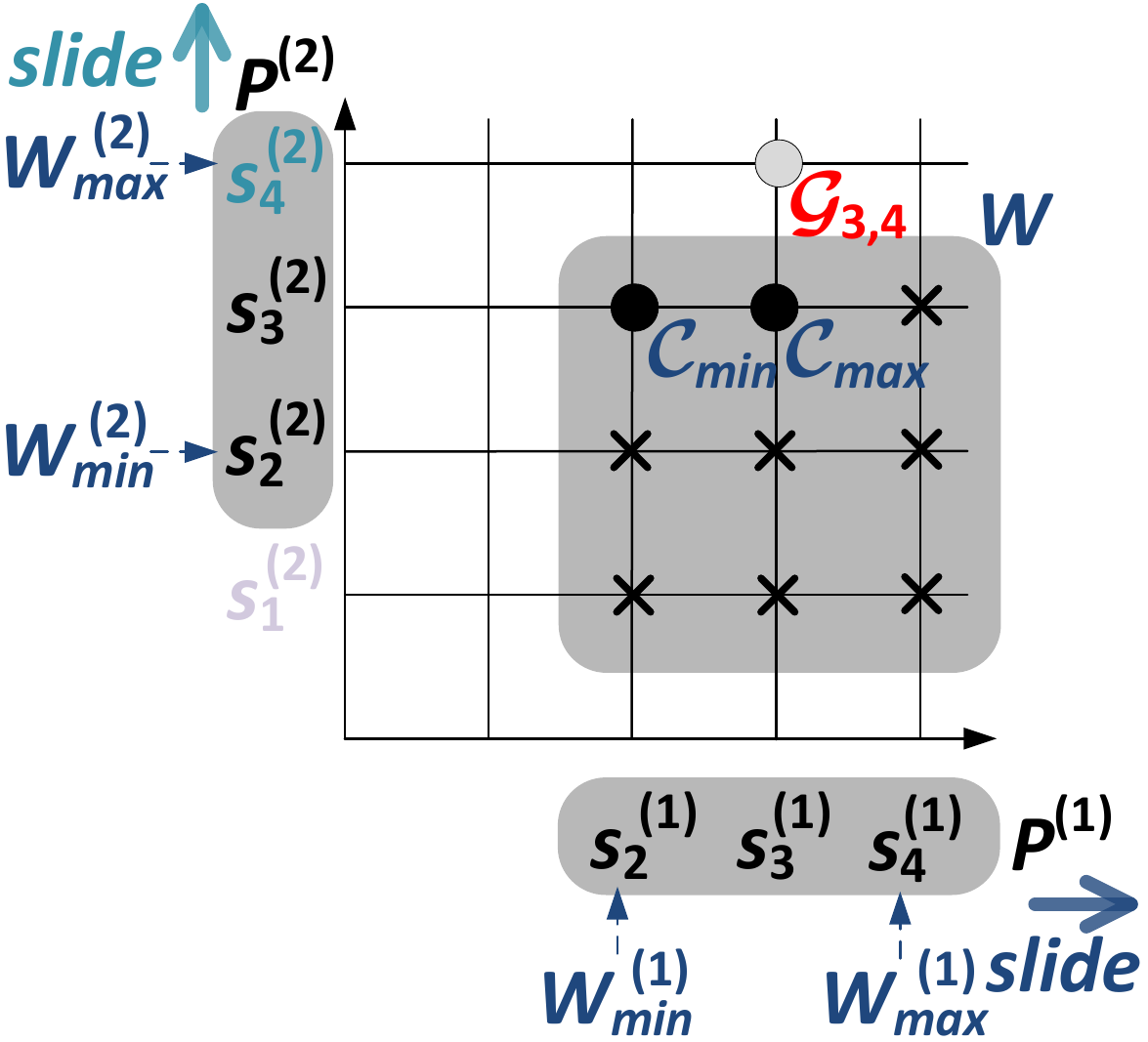}
        \end{minipage}}
    \caption{Restrictions and the slide of the $n$-dimensional window. Assume the arrival of local states is $s^{(1)}_0, s^{(1)}_1, s^{(1)}_2, s^{(2)}_0, s^{(2)}_1, s^{(2)}_2, s^{(1)}_3, s^{(2)}_3, s^{(1)}_4, s^{(2)}_4, \cdots$.}
    \label{F:examples}
\end{figure}

In this section, we discuss the update of {\it Lat-Win} when the $n$-dimensional window slides. Informally, the window slides as a new event is executed on $P^{(k)}$ and $P^{(k)}$ arrives at a new local state.

When a new local state from $P^{(k)}$ arrives, the stale local state (i.e., the old $W^{(k)}_{min}$) will be discarded. {\it Lat-Win} will grow with the CGSs containing the newly arrived local state, and prune the CGSs containing the stale local state, as shown in Fig. \ref{F:design overview}. Since the intersection between the set of new CGSs and the set of stale CGSs is empty, the growing and pruning of {\it Lat-Win} can be proceeded in any order. In this work, we first add newly obtained CGSs to {\it Lat-Win} and then prune the stale CGSs. During the growing and pruning process, $\mathcal{C}_{min}$ and $\mathcal{C}_{max}$ are also updated for further updates of {\it Lat-Win}.

We characterize the evolution of {\it Lat-Win} in three steps:
\begin{itemize}
    \item Lemma 3 defines the {\it restrictions} of lattice, which serves as the basis for further growing and pruning;
    \item Theorem 4 defines the condition when {\it Lat-Win} can grow and Theorem 5 identifies the new $\mathcal{C}_{min}$ and $\mathcal{C}_{max}$ when {\it Lat-Win} grows;
    \item Theorem 6 defines the condition when {\it Lat-Win} can prune and Theorem 7 identifies the new $\mathcal{C}_{min}$ and $\mathcal{C}_{max}$ when {\it Lat-Win} prunes.
\end{itemize}

\noindent The growing and pruning are dual, as well as the updates of $\mathcal{C}_{min}$ and $\mathcal{C}_{max}$.

\subsubsection{Restrictions}

Before we discuss the update of {\it Lat-Win}, we first introduce the notion of {\it restrictions}. When we obtain a global state and decide that it is not consistent, we can induce a specific region containing only inconsistent global states. The specific regions are also called {\it restrictions} in \cite{Mattern89}.

The geometric interpretation can be illustrated by the example in Fig. \ref{F:restrictions}, global states $\mathcal{G}_{1,2}$ = ($s^{(1)}_{1}, s^{(2)}_{2}$) and $\mathcal{G}_{4,5}$ = ($s^{(1)}_{4}, s^{(2)}_{5}$) are not consistent ($s^{(2)}_{2} \rightarrow s^{(1)}_{1}$ and $s^{(1)}_{4} \rightarrow s^{(2)}_{5}$ in Fig. \ref{F:space time diagram}). When looking from $\mathcal{C}_{0,2}$, $\mathcal{G}_{1,2}$ makes the lower gray region have no CGSs. When looking from $\mathcal{C}_{5,5}$, $\mathcal{G}_{4,5}$ makes the upper gray region have no CGSs. Formally, we have the following lemma: \\

\noindent{\bf Lemma 3.} To a CGS $\mathcal{C}$ of a lattice, and two global states $\mathcal{G}_1, \mathcal{G}_2$, $\mathcal{G}_1[i]$ ($\mathcal{G}_2[i]$) is the first (last) local state after (before) $\mathcal{C}[i]$ on $P^{(i)}$, $\forall k\neq i$, $\mathcal{G}_1[k] = \mathcal{G}_2[k] = \mathcal{C}[k]$,
\begin{itemize}
    \item[1.] If $\mathcal{G}_1$ is not CGS, then $\exists j\neq i, \mathcal{G}_1[j] \rightarrow \mathcal{G}_1[i]$, and none of the global states in the following set is CGS:
        \{ $\mathcal{G} | \mathcal{G}[j]\rightarrow \mathcal{G}_1[j]$ or $\mathcal{G}[j] = \mathcal{G}_1[j]$, $\mathcal{G}_1[i]\rightarrow \mathcal{G}[i]$ or $\mathcal{G}_1[i] = \mathcal{G}[i]$\};
    \item[2.] If $\mathcal{G}_2$ is not CGS, then $\exists j\neq i, \mathcal{G}_2[i] \rightarrow \mathcal{G}_2[j]$, and none of the global states in the following set is CGS:
        \{ $\mathcal{G} | \mathcal{G}_2[j]\rightarrow \mathcal{G}[j]$ or $\mathcal{G}_2[j] = \mathcal{G}[j]$, $\mathcal{G}[i]\rightarrow \mathcal{G}_2[i]$ or $\mathcal{G}[i] = \mathcal{G}_2[i]$\}.
\end{itemize}

\noindent{\it Proof:}

3.1: If $\mathcal{G}_1$ is not CGS, it is easy to verify that $\exists j\neq i, \mathcal{G}_1[j] \rightarrow \mathcal{G}_1[i]$. To any global state $\mathcal{G}$ in $\{\mathcal{G} | \mathcal{G}[j]\rightarrow \mathcal{G}_1[j]$ or $\mathcal{G}[j] = \mathcal{G}_1[j], \mathcal{G}_1[i]\rightarrow \mathcal{G}[i]$ or $\mathcal{G}_1[i] = \mathcal{G}[i]\}$, it is easy to verify that $\mathcal{G}[j] \rightarrow \mathcal{G}[i]$. Thus, global state $\mathcal{G}$ is not consistent, and none of the global states in the set is CGS.

3.2: The proof is dual as above. \qed\\

\subsubsection{Growing of {\it Lat-Win}}

On the arrival of a new local state $s^{(k)}_{i}$, the $n$-dimensional window slides in $P^{(k)}$'s dimension, i.e., $W^{(k)}_{max} = s^{(k)}_{i}$, and a set of newly obtained CGSs (containing $s^{(k)}_{i}$) will be added into {\it Lat-Win}. We find that the growing process does not have to explore the whole combinational space of the new local state with all local states from every $W^{(k)}$. If {\it Lat-Win} is not empty, it will grow from $\mathcal{C}_{max}$ in {\it Lat-Win}. The reason is that, if the next global state growing from $\mathcal{C}_{max}$ is not consistent, as $\mathcal{G}_1$ in Lemma 3, it can be proved that the global states containing the newly arrived local state as a constituent are all in the restriction induced by a further global state and therefore not consistent. When {\it Lat-Win} is empty, the lattice can grow iff one CGS can be obtained containing the new local state and a lower bound of some local window. This is because as discussed in Theorem 2, the new $\mathcal{C}_{min}$ should contain at least a lower bound of a local window. Formally, \\

\noindent{\bf Theorem 4.} When a new event $e^{(k)}_{i}$ is executed on $P^{(k)}$ and the new local state $s^{(k)}_{i}$ from $P^{(k)}$ arrives,
\begin{itemize}
    \item[1.] If {\it Lat-Win} $\neq \emptyset$, then {\it Lat-Win} can grow iff $\mathcal{C}_{max}[k] = s^{(k)}_{i-1}$ (the old $W^{(k)}_{max}$) and global state $\mathcal{G}$ ($\mathcal{G}[k] = s^{(k)}_{i}$, $\forall j \neq k, \mathcal{G}[j] = \mathcal{C}_{max}[j]$) is CGS;
    \item[2.] If {\it Lat-Win} = $\emptyset$, then {\it Lat-Win} can grow iff $\{\mathcal{C}|\mathcal{C}[k] = s^{(k)}_i$, $\exists j\neq k, \mathcal{C}[j] = W^{(j)}_{min}, \mathcal{C}$ is CGS\}$\neq \emptyset$.
\end{itemize}

\noindent{\it Proof:}

4.1: ``$\Rightarrow$'': We first prove that if {\it Lat-Win} is not empty and can grow, $\mathcal{C}_{max}[k] = s^{(k)}_{i-1}$. We prove it by contradiction. If $\mathcal{C}_{max}[k] \neq s^{(k)}_{i-1}$, $\mathcal{C}_{max}[k] \rightarrow s^{(k)}_{i-1}$ and $\exists S_{succ}(\mathcal{C}_{max}[k]) \in W^{(i)}$. From the proof of Theorem 2.1, we can easy to verify that $\exists j\neq k, \mathcal{C}_{max}[j] = W^{(j)}_{max}$ and $W^{(j)}_{max} \rightarrow S_{succ}(\mathcal{C}_{max}[k])$. Note that $S_{succ}(\mathcal{C}_{max}[k]) \rightarrow s^{(k)}_{i}$. Thus, to any new global states $\mathcal{G}'$ containing $s^{(k)}_{i}$, $\mathcal{G}'[j] \rightarrow \mathcal{G}'[k]$. By Lemma 3, {\it Lat-Win} cannot grow, which is contract to that {\it Lat-Win} can grow. Thus, if {\it Lat-Win} is not empty and can grow, $\mathcal{C}_{max}[k] = s^{(k)}_{i-1}$.

We then prove that if {\it Lat-Win} is not empty and can grow, global state $\mathcal{G}$ ($\mathcal{G}[k] = s^{(k)}_{i}$, $\forall j \neq k, \mathcal{G}[j] = \mathcal{C}_{max}[j]$) is CGS. We prove it by contradiction. If $\mathcal{G}$ is not CGS, $\exists j\neq k, \mathcal{G}[j]\rightarrow \mathcal{G}[k]$, $E_{succ}(\mathcal{C}_{max}[j]) \rightarrow E_{succ}(\mathcal{C}_{max}[k])$ ($\mathcal{G}[j] = \mathcal{C}_{max}[j], \mathcal{G}[k] = succ(\mathcal{C}_{max}[k])$). If $\mathcal{C}_{max}[j] \neq W^{(j)}_{max}$, by the proof of Theorem 2.1, we can easily verify that $\exists m\not\in \{j, k\}, \mathcal{C}_{max}[m] = W^{(m)}_{max}$, $W^{(m)}_{max}\rightarrow S_{succ}(\mathcal{C}_{max}[j])$ and $E_{succ}(W^{(m)}_{max}) \rightarrow E_{succ}(\mathcal{C}_{max}[j])$. Thus, $E_{succ}(W^{(m)}_{max}) \rightarrow E_{succ}(\mathcal{C}_{max}[k])$. Thus, to any new global states $\mathcal{G}'$ containing $s^{(k)}_{i}$, $\mathcal{G}'[m] \rightarrow \mathcal{G}'[k]$. By Lemma 3, {\it Lat-Win} cannot grow. Thus, if {\it Lat-Win} is not empty and can grow, global state $\mathcal{G}$ ($\mathcal{G}[k] = s^{(k)}_{i}$, $\forall j \neq k, \mathcal{G}[j] = \mathcal{C}_{max}[j]$) is CGS.

``$\Leftarrow$'': Global state $\mathcal{G}$ is CGS, thus {\it Lat-Win} can grow.

4.2: ``$\Rightarrow$'': It is easy to verify the theorem by Theorem 2.2. It can be proved by contradiction. If $\{\mathcal{C}|\mathcal{C}[k] = s^{(k)}_i$, $\exists j\neq k, \mathcal{C}[j] = W^{(j)}_{min}, \mathcal{C}$ is CGS\}$ = \emptyset$, after the process of growing, the new $\mathcal{C}_{min}$ contains $s^{(k)}_i$, and $\forall j\neq k, W^{(j)}_{min} \rightarrow \mathcal{C}_{min}[j]$. By combining the predecessor local state of each $\mathcal{C}_{min}[j]$ with $\mathcal{C}_{min}$, a violation of irreflexivity will be inferred, dual as the proof of Theorem 2.1.

``$\Leftarrow$'': $\{\mathcal{C}|\mathcal{C}[k] = s^{(k)}_i$, $\exists j\neq k, \mathcal{C}[j] = W^{(j)}_{min}, \mathcal{C}$ is CGS\}$\neq \emptyset$, then {\it Lat-Win} can grow. \qed \\

We illustrate the theorem by three examples in Fig. \ref{F:example 1}, Fig. \ref{F:example 2}, and Fig. \ref{F:example 3}, on the arrival of $s^{(2)}_3$, $s^{(1)}_4$, and $s^{(2)}_4$, respectively. In Fig. \ref{F:example 1}, the current {\it Lat-Win} is empty and $s^{(2)}_{3}$ arrives. The lattice can grow iff the global state $\mathcal{G}_{1,3}$ = ($s^{(1)}_{1}$, $s^{(2)}_{3}$) is CGS. Note that $\mathcal{G}_{1,3}$ is CGS (as shown in Fig. \ref{F:restrictions}). Thus {\it Lat-Win} can grow to the new lattice in Fig. \ref{F:example 2}. In Fig. \ref{F:example 2}, the current {\it Lat-Win} is not empty and $s^{(1)}_{4}$ arrives. {\it Lat-Win} can grow iff $\mathcal{C}_{max}[1] = s^{(1)}_{3}$ and the global state $\mathcal{G}_{4,3}$ = ($s^{(1)}_{4}$, $s^{(2)}_{3}$) is CGS. Note that $\mathcal{G}_{4,3}$ is not CGS (in Fig. \ref{F:restrictions}). Thus {\it Lat-Win} cannot grow, as shown in Fig. \ref{F:example 3}. In Fig. \ref{F:example 3}, the current {\it Lat-Win} is not empty and $s^{(2)}_{4}$ arrives. {\it Lat-Win} can grow iff $\mathcal{C}_{max}[2] = s^{(2)}_{3}$ and the global state $\mathcal{G}_{3,4}$ = ($s^{(1)}_{3}$, $s^{(2)}_{4}$) is CGS. Note that $\mathcal{G}_{3,4}$ is CGS (in Fig. \ref{F:restrictions}). Thus {\it Lat-Win} can grow to the new lattice in Fig. \ref{F:design overview 1}.

The maximal and minimal CGSs are important to the update of {\it Lat-Win}. Thus, we discuss how to locate $C_{max}$ and $C_{min}$ after the growing of {\it Lat-Win} for further updates. After the growing of {\it Lat-Win}, the new $C_{max}$ should contain the new local state as a constituent. If {\it Lat-Win} was empty and grows with the new local state, $C_{min}$ should contain the new local state as a constituent.

For example in Fig. \ref{F:example 1}, {\it Lat-Win} is empty and can grow with the newly arrived local state $s^{(2)}_{3}$, the new $\mathcal{C}_{max}[2]$ = $s^{(2)}_{3}$ and the new $\mathcal{C}_{min}[2]$ = $s^{(2)}_{3}$, as shown in Fig. \ref{F:example 2}. Formally, \\

\noindent{\bf Theorem 5.} When a new event $e^{(k)}_{i}$ is executed on $P^{(k)}$ and the new local state $s^{(k)}_{i}$ from $P^{(k)}$ arrives,
\begin{itemize}
    \item[1.] If {\it Lat-Win} can grow, then $\mathcal{C}_{max}[k] = s^{(k)}_{i}$ (the new $W^{(k)}_{max}$); else $\mathcal{C}_{max}$ remains.
    \item[2.] If {\it Lat-Win} = $\emptyset$ and can grow, then $\mathcal{C}_{min}[k] = s^{(k)}_{i}$; else $\mathcal{C}_{min}$ remains.
\end{itemize}

\noindent{\it Proof:}

5.1: If {\it Lat-Win} can grow, all the new CGSs contain $s^{(k)}_i$ as a constituent. The CGS $\mathcal{G}$ in Theorem 4.1 ensures that the new maximal CGS is at least ``larger than'' $\mathcal{G}$. Thus, the new $\mathcal{C}_{max}$ is in the set of the new CGSs, and $\mathcal{C}_{max}[k] = s^{(k)}_{i}$.

5.2: If {\it Lat-Win} = $\emptyset$ and can grow, it is easy to verify $\mathcal{C}_{min}[k] = s^{(k)}_{i}$. \qed \\

\subsubsection{Pruning of {\it Lat-Win}}

On the arrival of a new local state, after the growing of new CGSs, {\it Lat-Win} will prune the CGSs which contain the stale local state. The pruning does not have to explore the whole lattice to check whether a CGS contains the stale local state. Intuitively, {\it Lat-Win} can prune, iff {\it Lat-Win} is not empty and $\mathcal{C}_{min}$ contains the stale local state. Formally, \\

\noindent{\bf Theorem 6.} When a new event $e^{(k)}_{i}$ is executed on $P^{(k)}$ and the new local state $s^{(k)}_{i}$ from $P^{(k)}$ arrives, after the growing, {\it Lat-Win} can prune, iff {\it Lat-Win} $\neq\emptyset$ and $\mathcal{C}_{min}[k] \rightarrow W^{(k)}_{min}$.

\noindent{\it Proof:}

``$\Rightarrow$'': If {\it Lat-Win} can prune, that is, there is at least a CGS containing the stale local state (the old $W^{(k)}_{min}$). Thus, $\mathcal{C}_{min}[k]$ equals the old $W^{(k)}_{min}$, and $\mathcal{C}_{min}[k] \rightarrow W^{(k)}_{min}$.

``$\Leftarrow$'': If {\it Lat-Win} $\neq \emptyset$ and $\mathcal{C}_{min}[k] \rightarrow W^{(k)}_{min}$, $\mathcal{C}_{min}[k]$ contains the stale local state. Thus, {\it Lat-Win} can prune. \qed \\

For example, in Fig. \ref{F:example 2}, on the arrival of $s^{(1)}_4$, $\mathcal{C}_{min}[1] = s^{(1)}_1$ and $\mathcal{C}_{min}[1]\rightarrow W^{(1)}_{min}$. Thus {\it Lat-Win} can prune, as shown in Fig. \ref{F:example 3}. In Fig. \ref{F:example 3}, on the arrival of $s^{(2)}_4$, $\mathcal{C}_{min}[2] = s^{(2)}_3$ and $\mathcal{C}_{min}[2]\not\rightarrow W^{(2)}_{min}$. Thus {\it Lat-Win} does not have to prune, as shown in Fig. \ref{F:design overview 1}.

We then discuss how to locate $C_{max}$ and $C_{min}$ after the pruning of {\it Lat-Win} for further updates. When a new local state from $P^{(k)}$ arrives, after the pruning of {\it Lat-Win}, if {\it Lat-Win} prunes to be empty, the maximal and minimal CGSs are {\it null}. If {\it Lat-Win} prunes to be non-empty, the minimal CGS should contain the new $W^{(k)}_{min}$. Formally, \\

\noindent{\bf Theorem 7.} When a new event $e^{(k)}_{i}$ is executed on $P^{(k)}$ and the new local state $s^{(k)}_{i}$ from $P^{(k)}$ arrives, after the growing,
\begin{itemize}
    \item[1.] If $\mathcal{C}_{max}[k]\rightarrow W^{(k)}_{min}$, then $\mathcal{C}_{max} = null$; else $\mathcal{C}_{max}$ remains.
    \item[2.] If $\mathcal{C}_{max}[k]\rightarrow W^{(k)}_{min}$, then $\mathcal{C}_{min} = null$; else if {\it Lat-Win} can prune, then $\mathcal{C}_{min}[k] = W^{(k)}_{min}$; else $\mathcal{C}_{min}$ remains.
\end{itemize}

\noindent{\it Proof:}

7.1: If $\mathcal{C}_{max}[k]\rightarrow W^{(k)}_{min}$, all CGSs in {\it Lat-Win} contain the stale local state. Thus, {\it Lat-Win} prunes to be empty, and $\mathcal{C}_{max} = null$.

7.2: If $\mathcal{C}_{max}[k]\rightarrow W^{(k)}_{min}$, {\it Lat-Win} prunes to be empty, and $\mathcal{C}_{min} = null$; If $\mathcal{C}_{max}[k]\not\rightarrow W^{(k)}_{min}$ and {\it Lat-Win} can prune, there is at least a CGS containing $W^{(k)}_{min}$. Thus, the new $\mathcal{C}_{min}[k] = W^{(k)}_{min}$. \qed \\

For example, in Fig. \ref{F:example 2}, on the arrival of $s^{(1)}_4$, {\it Lat-Win} can prune and $\mathcal{C}_{max}[1]\not\rightarrow W^{(1)}_{min}$, then the new $\mathcal{C}_{min}[1] = W^{(1)}_{min}$ and $\mathcal{C}_{max}$ is not changed, as shown in Fig. \ref{F:example 3}.

\section{Lat-Win - Online Maintenance Algorithm}
\label{sec:algorithm}

In this section, we present the design of the {\it Lat-Win} maintenance algorithm, based on the theoretical characterization above. $\mathcal{C}_{min}$ and $\mathcal{C}_{max}$ serve as two anchors in maintaining {\it Lat-Win}. When a new local state arrives, {\it Lat-Win} grows from $\mathcal{C}_{max}$ and prunes from $\mathcal{C}_{min}$. After the growing and pruning, $\mathcal{C}_{min}$ and $\mathcal{C}_{max}$ are also updated for further maintenance of {\it Lat-Win}.

$P_{che}$ is in charge of collecting and processing the local states sent from non-checker processes. Upon initialization, $P_{che}$ gets the window size $w$ and initializes $n$ local windows $W^{(k)}$ over local states from each $P^{(k)}$. Upon receiving a new local state from $P^{(k)}$, $P_{che}$ first enqueues the local state into $Que^{(k)}$ and then updates {\it Lat-Win} in the order of growing and pruning. Pseudo codes of the maintenance algorithm are listed in Algorithm 1.

\begin{algorithm}[htbp]
\SetAlgoVlined
\textbf{Upon} Initialization\\
    get window size $w$\;
    initialize window buffers $W^{(k)}$\;
\textbf{Upon} Receiving local state ($s^{(k)}_i$) from $P^{(k)}$\\
    $Que^{(k)}.enque(s^{(k)}_i)$\;
    \If{$s^{(k)}_i = Que^{(k)}.head()$}
    {
        pop the front continuous local states of $Que^{(k)}$ to the end of $InputQue$\;
        trigger $update()$\;
    }
    \vspace{0.3cm}
    $update()$\\
    \While{$InputQue\neq\emptyset$}
    {
        pop $s^{(k)}_i = InputQue.head()$\;
        push $s^{(k)}_i$ into $W^{(k)}$ \tcc*[r]{$W^{(k)}_{max} = s^{(k)}_i$}
        $grow\_lattice(s^{(k)}_i, k)$ \tcc*[r]{Algorithm 2}
        $prune\_lattice(\mathcal{C}_{min},k)$ \tcc*[r]{Algorithm 3}
    }
\caption{{\it Lat-Win} maintenance algorithm\label{A:Checker-Process}}
\end{algorithm}

\subsection{Growing of {\it Lat-Win}}

On the arrival of a new local state, the process of growing consists of three steps. First, it is checked whether {\it Lat-Win} can grow, as discussed in Theorem 4. If yes, {\it Lat-Win} will grow with a set of new CGSs containing the new arrived local state. During the step of growing, $\mathcal{C}_{max}$ and $\mathcal{C}_{min}$ are updated too, as discussed in Theorem 5. Pseudo codes of growing are listed in Algorithm 2.

When {\it Lat-Win} is empty, the lattice can grow iff the set $S$ in line 2 contains a CGS, as discussed in Theorem 4.2. If $S$ has a CGS, the $grow()$ sub-routine will be triggered to add the new CGSs. When {\it Lat-Win} is not empty, the lattice can grow iff $\mathcal{C}_{max}$ and the next global state satisfy the condition defined in Theorem 4.1, as shown in line 5-8. If the condition is satisfied, the $grow()$ sub-routine will be triggered to add the new CGSs. The growing of {\it Lat-Win} is achieved by recursively adding all the predecessors and successors of a CGS, as shown in line 10-19. During the growing process, $\mathcal{C}_{max}$ and $\mathcal{C}_{min}$ are also updated. Theorem 5 ensures that $\mathcal{C}_{max}$ can be found in the new added CGSs, and that when {\it Lat-Win} was empty, $\mathcal{C}_{min}$ can be found in the new added CGSs, as shown in line 12-13.

\begin{algorithm}[htbp]
\SetAlgoVlined
\uIf{{\it Lat-Win} = $\emptyset$}
{
    $S$ = $\{\mathcal{C}|\mathcal{C}[k]$ = $s^{(k)}_i$, $\exists j\neq k$, $\mathcal{C}[j]$ = $W^{(j)}_{min}$, $\mathcal{C}$ is CGS\}\;
    \If{$S\neq \emptyset$} {
        get a CGS $\mathcal{C}$ from $S$;\hspace{0.05in}$grow(\mathcal{C}$);

    }
}
\ElseIf{$\mathcal{C}_{max}[k] = s^{(k)}_{i-1}$}
{
    combine $\mathcal{C}_{max}$ and $s^{(k)}_{i}$ to get a global state $\mathcal{G}$\;
    \If{$\mathcal{G}$ is CGS}
    {
        connect $\mathcal{G}$ to $\mathcal{C}_{max}$;\hspace{0.05in}$grow(\mathcal{G})$\;
    }
}
\vspace{0.3cm}
$grow(\mathcal{C}$)\\
Set $prec(\mathcal{C}) = \{\mathcal{C}'|\forall i, \mathcal{C}'[i]\in W^{(i)}, \mathcal{C}'$ is CGS, $\mathcal{C}'\prec \mathcal{C}$\}\;
Set $sub(\mathcal{C}) = \{\mathcal{C}'|\forall i, \mathcal{C}'[i]\in W^{(i)}, \mathcal{C}'$ is CGS, $\mathcal{C}\prec \mathcal{C}'$\}\;
\lIf{$prec(\mathcal{C}) = \emptyset$}
{
    $\mathcal{C}_{min} = \mathcal{C}$\;
}
\lIf{$sub(\mathcal{C}) = \emptyset$}
{
    $\mathcal{C}_{max} = \mathcal{C}$\;
}
\ForEach{$\mathcal{C}'$ in $prec(\mathcal{C})$}
{
    \If{$\mathcal{C}'$ does not exist}
    {
        connect $\mathcal{C}'$ to $\mathcal{C}$;\hspace{0.05in}$grow(\mathcal{C}')$\;
    }
}
\ForEach{$\mathcal{C}'$ in $sub(\mathcal{C})$}
{
    \If{$\mathcal{C}'$ does not exist}
    {
        connect $\mathcal{C}$ to $\mathcal{C}'$;\hspace{0.05in}$grow(\mathcal{C}')$\;
    }
}
\caption{$grow\_lattice(s^{(k)}_i, k)$\label{A:grow_lattice($s^{(k)}_i, k$)}}
\end{algorithm}

\subsection{Pruning of {\it Lat-Win}}

Dually, the process of pruning consists of three steps as well. First, it is checked whether {\it Lat-Win} can prune, as discussed in Theorem 6. If yes, {\it Lat-Win} will prune the set of CGSs which contain the stale local state. During the step of pruning, $\mathcal{C}_{max}$ and $\mathcal{C}_{min}$ are updated too, as discussed in Theorem 7. Pseudo codes of pruning are listed in Algorithm 3.

The lattice can prune iff the condition in line 1 is satisfied, as discussed in Theorem 6. If $\mathcal{C}_{max}[k]$ is the stale local state ($\mathcal{C}[k]$ in line 2), {\it Lat-Win} will prune to be empty. Otherwise, the CGSs which contain the stale local state will be deleted, as shown in line 5-16. During the pruning process, $\mathcal{C}_{min}$ is also updated. Theorem 7 ensures that $\mathcal{C}_{min}$ can be found in the CGSs which contain the new $W^{(k)}_{min}$, as shown in line 13-15.

\begin{algorithm}[htbp]
\SetAlgoVlined
\If{{\it Lat-Win} $\neq \emptyset$ \&\& $\mathcal{C}_{min}[k]\rightarrow W^{(k)}_{min}$}
{
    \eIf{$\mathcal{C}_{max}[k] = \mathcal{C}[k]$}
    {
        {\it Lat-Win} = $\emptyset$, $\mathcal{C}_{max} = \mathcal{C}_{min} = null$\;
    }
    {
        Set $S = \{\mathcal{C}\}$\;
        \While{$S\neq\emptyset$}
        {
            pop $\mathcal{C}'$ from $S$\;
            Set $sub(\mathcal{C}') = \{\mathcal{C}''|\mathcal{C}'\prec\mathcal{C}'', \mathcal{C}''$ is CGS\}\;
            \ForEach{$\mathcal{C}''$ in $sub(\mathcal{C}')$}
            {
                delete the connection between $\mathcal{C}'$ and $\mathcal{C}''$\;
                \uIf{$\mathcal{C}''[k] = \mathcal{C}'[k]$ \&\& $\mathcal{C}''\not\in S$}
                {
                    add $\mathcal{C}''$ into $S$\;
                }
                \ElseIf{$\mathcal{C}''[k] = W^{(k)}_{min}$}
                {
                    Set $prec(\mathcal{C}'') = \{\mathcal{C}'''|\mathcal{C}'''\prec\mathcal{C}'', \mathcal{C}'''$ is CGS\} \tcc*[r]{without $\mathcal{C}'$}
                    \lIf{$prec(\mathcal{C}'')= \emptyset$}
                    {
                        $\mathcal{C}_{min} = \mathcal{C}''$\;
                    }
                }
            }
            delete $\mathcal{C}'$\;
        }
    }
}
\caption{$prune\_lattice(\mathcal{C},k)$\label{A:prune_lattice(C,k)}}
\end{algorithm}

\subsection{Complexity Analysis}
\label{sec:complexity analysis}

Regarding the space for storing a single CGS as one unit, the worst-case space cost of the original lattice maintenance is $O(p^n)$, where $p$ is the upper bound of number of events of each non-checker process, and $n$ is the number of non-checker processes. However, the worst-case space cost of sliding windows over the original lattice is bounded by the size $w$ of the windows, that is, $O(w^n)$, where $w$ is a fixed number and much less than $p$. Due to the incremental nature of Algorithm 2, the space cost of the incremental part of {\it Lat-Win} in each time of growing is $O(w^{n-1})$.

The worst-case time cost of growing (Algorithm \ref{A:grow_lattice($s^{(k)}_i, k$)}) happens when all the global states in the blue rectangle in Fig. \ref{F:design overview 2} are CGSs. Thus, the worst-case time of growing is $O(n^3w^{n-1})$, where $w^{n-1}$ is the number of the global states in the blue rectangle and $n^3$ is the time cost of checking whether a global state is consistent.

The worst-case time cost of pruning (Algorithm \ref{A:prune_lattice(C,k)}) happens when all the CGSs in the left shaded rectangle in Fig. \ref{F:design overview 3} should be discarded. Thus, the worst-case time of pruning is $O(w^{n-1})$, where $w^{n-1}$ is the worst-case number of the CGSs in the left shaded rectangle.

From the performance analysis we can see that, by tuning the sliding window size $w$, the cost of asynchronous event stream processing can be effectively bounded. This justifies the adoption of sliding windows when only recent part of the event streams are needed by the tracking/monitoring application, and the application needs to strictly bound the cost of event processing.

\section{Experimental Evaluation}
\label{sec:evaluation}

In this section, we further investigate the performance of {\it Lat-Win} via a case study. We first describe a smart office scenario to demonstrate how our approach supports context-awareness in asynchronous pervasive computing scenarios. Then, we describe the experiment design. Finally, we discuss the evaluation results.

\subsection{Achieving Context-awareness by On-line Processing of Asynchronous Event Streams}

We simulate a smart office scenario, where a context-aware application on Bob's mobile phone automatically turns the phone to silent mode when Bob attends a lecture\cite{Huang11}.

The application detects that Bob attends a lecture by specification of the concurrency property: $C_{1}$: {\it location of Bob is the meeting room, a speaker is in the room, and a presentation is going on} \cite{Huang11}. The application needs to turn the phone to silent mode when the property definitely holds. Formally, $C_{1} = Def(\phi_{1}\wedge\phi_{2}\wedge\phi_{3})$, which is explained in detail below.

The location context is detected by Bob's smart phone. When the phone connects to the access point in the meeting room, we assume that Bob is in this room. Specifically, non-checker process $P^{(1)}$ is deployed on Bob's smart phone, which updates the phone's connection to access points. Local predicate $\phi_{1}$ = {\it``the user's smart phone connects to the AP inside the meeting room''} is specified over $P^{(1)}$.

An RFID reader is deployed in the room to detect the speaker. Specifically, non-checker process $P^{(2)}$ is deployed on the RFID reader, and local predicate $\phi_{2}$ = {\it``the RFID reader detects a speaker''} is specified over $P^{(2)}$.

We detect that a presentation is going on if the projector is working. Specifically, non-checker process $P^{(3)}$ is deployed over the projector, and local predicate $\phi_{3}$ = {\it``the projector is on''} is specified over $P^{(3)}$.

Observe that the mobile phone, the RFID reader, and the projector do not necessarily have synchronized clocks, and Bob may not be willing to synchronize his mobile with other devices due to privacy concerns. They suffer from message delay of wireless communications. Furthermore, the recent data is more informative and useful to the context-aware application than stale data, thus the sliding window can be imposed over the streams of context.

Non-checker processes produce event streams and communications among them help establish the happen-before relation among events. A checker process $P_{che}$ is deployed on the context-aware middleware to collect local states with logical clock timestamps from non-checker processes, and maintain {\it Lat-Win} at runtime. Based on {\it Lat-Win}, $P_{che}$ can further detect the concurrency property $C_{1}$ \cite{Huang11} and notify the mobile phone to turn to silent mode.

The detection of concurrency properties assumes the availability of an underlying context-aware middleware. We have implemented the middleware based on one of our research projects - {\it Middleware Infrastructure for Predicate detection in Asynchronous environments} (MIPA) \cite{MIPA, Huang10b, Huang11}.

\subsection{Experiment Design}

The user's and speaker's stay inside and outside of the meeting room, as well as the working period of the projector are generated following the Poisson process. Specifically, the sensors collect context data every 1 min. We model the message delay by exponential distribution. The average duration of local contextual activities is 25 mins and the interval between contextual activities is 5 mins. Lifetime of the experiment is up to 100 hours.

In the experiments, we first study the benefits of imposing sliding window over asynchronous event streams in the percentage of detection $Perc_{det}$ and the percentage of space cost $Perc_{s}$. $Perc_{det}$ is defined as the ratio of $\frac{N_{Lat-Win}}{N_{LAT}}$. Here, $N_{Lat-Win}$ denotes the number of times the algorithm detects the specified property on {\it Lat-Win}. $N_{LAT}$ denotes the number of times the algorithm detects the specified property on the original lattice {\it LAT}. $Perc_{s}$ is defined as the ratio of $\frac{S_{Lat-Win}}{S_{LAT}}$. Here, $S_{Lat-Win}$ denotes the average size of {\it Lat-Win} as the window slides. $S_{LAT}$ denotes the size of the original lattice over the lifetime of the experiment.

Then, we study the performance of {\it Lat-Win} in the probability of detection $Prob_{det}$, the space cost $S_{Lat-Win}$ and time cost $T_{Lat-Win}$. Here, $Prob_{det}$ is defined as the ratio of $\frac{N_{physical}}{N_{Lat-Win}}$. $N_{Lat-Win}$ is defined above and $N_{physical}$ denotes the number of times such property holds in the window when global time is available. $S_{Lat-Win}$ is defined above and $T_{Lat-Win}$ denotes the average time from the instant when $P_{che}$ is triggered to that when the detection finishes.

\subsection{Evaluation Results}

In this section, we first discuss the benefits of imposing sliding window over asynchronous event streams. Then we discuss the performance of the {\it Lat-Win} maintenance algorithm.

\subsubsection{Benefits of Sliding Window}

In this experiment, we investigate the impact of window size $w$ on the percentage of detection $Perc_{det}$ and the percentage of space cost $Perc_{s}$. We fix the average message delay to 0.5 s.

\begin{figure}[htbp]
\begin{center}
  \includegraphics[width=3.5in]{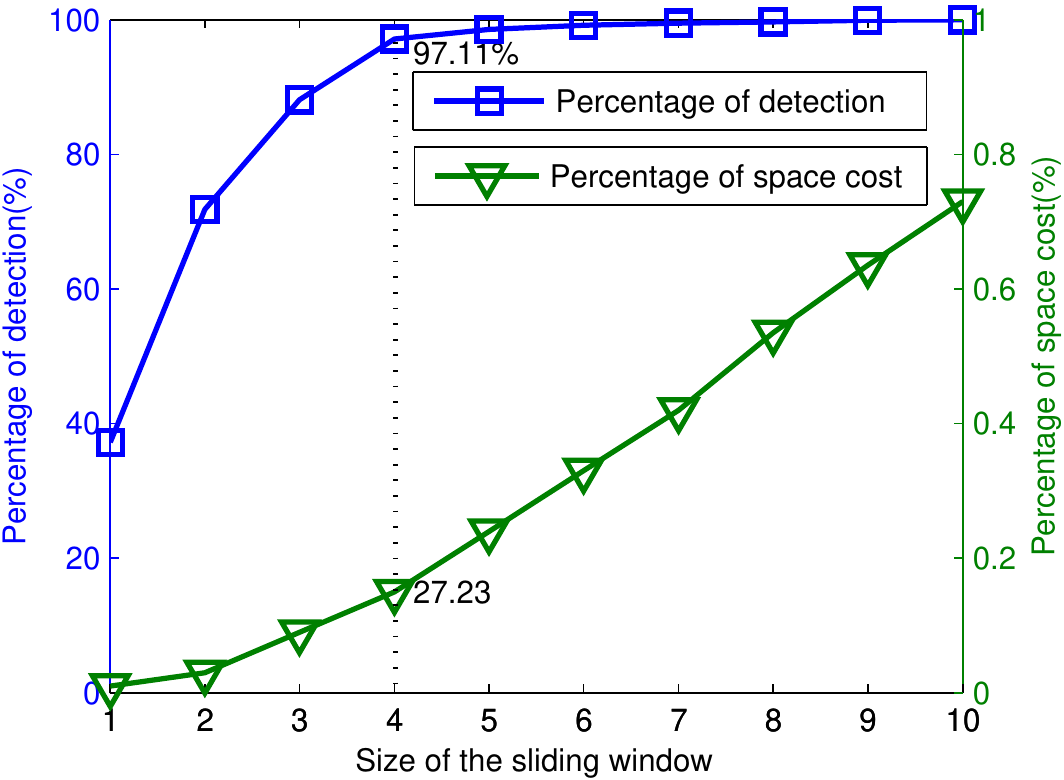}\\
  \centering\parbox[c]{2.5in}{\caption{Benefits of sliding window.}}
  \label{F:trade-off}
\end{center}
\end{figure}

The experiment shows that the sliding window enables the trade-off between $Perc_{det}$ and $Perc_{s}$. As shown in Fig. \ref{F:trade-off}, the increase of $w$ results in monotonic increase in both $Perc_{det}$ and $Perc_{s}$. When we tune $w$ from 1 to 4, $Perc_{det}$ (the upper blue line) increases quickly up to 97.11\%. When we tune $w$ from 4 to 10, $Perc_{det}$ increases slowly and remains quite high towards 100\%. $Perc_{s}$ (the lower green line) increases almost linearly as $w$ increases. When we set $w$ to 10, $Perc_{s}$ remains small than 1\%. We can find that, rather than maintaining the whole original lattice, imposing a quite small sliding window over asynchronous event streams can keep $Perc_{det}$ high and $Perc_{s}$ quite small. It indicates that recent data is more relevant and important to the application. When the window size is 4, the relative growth rate between $Perc_{det}$ and $Perc_{s}$ slows down, and $Perc_{det}$ is quite high, $S_{Lat-Win}$ is 27.23.  Thus, in the following experiments, we set the window size $w$ as 4 to study the performance of {\it Lat-Win}. (Notice that in different scenarios, the turning point of $w = 4$ may be different.)

\subsubsection{Performance of {\it Lat-Win}}

In this experiment, we investigate the performance of {\it Lat-Win} in the probability of detection $Prob_{det}$, the average space cost $S_{Lat-Win}$, and the average time cost $T_{Lat-Win}$. We first vary the asynchrony, i.e., the message delay. We also vary the window size $w$. Finally, we vary the number of non-checker processes $n$.

\paragraph{Effects of Tuning the Message Delay}

\begin{figure}[htbp]
\begin{center}
  \includegraphics[width=3.5in]{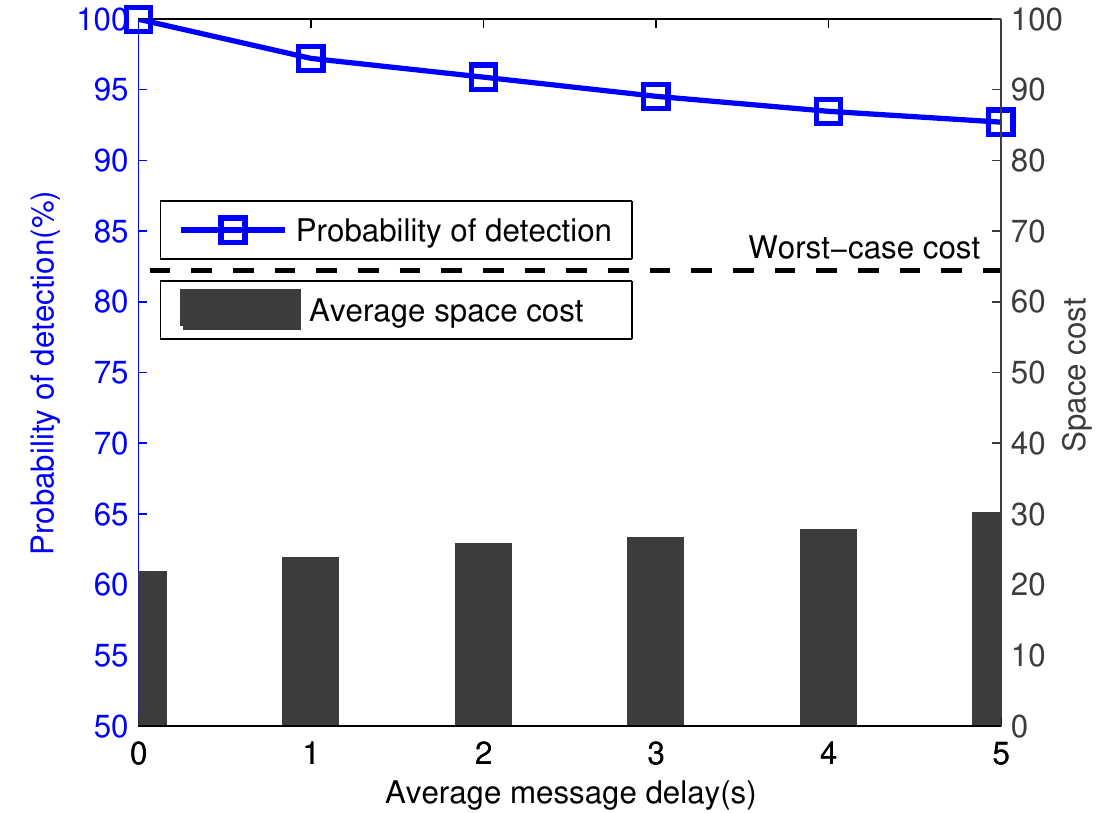}\\
  \centering\parbox[c]{2.5in}{\caption{Tuning the message delay.}}
  \label{F:message-delay}
\end{center}
\end{figure}

In this experiment, we study how the message delay affects the performance of {\it Lat-Win}. We fix the window size $w$ to 4. We find that when encountered with reasonably long message delay (less than 5 s), $Prob_{det}$ (the upper blue line) decreases slowly and remains over 85\%, as shown in Fig. \ref{F:message-delay}. When we tune the average message delay from 0 s to 5 s, $S_{Lat-Win}$ increases slowly to about 30, whereas the worst-case cost is $w^{n} = 64$, as discussed in Section \ref{sec:complexity analysis}. The decrease of $Prob_{det}$ and the increase of $S_{Lat-Win}$ are mainly due to the increase of the uncertainty caused by the asynchrony (i.e., the message delay). $N_{physical}$ is smaller than $N_{Lat-Win}$, because the detection algorithm may detect the property to be true but in physical world the property is not satisfied due to the increasing uncertainty caused by the asynchrony. Furthermore, the increase of message delay results in a slow increase in $T_{Lat-Win}$, which remains less than 1 ms.

\paragraph{Effects of Tuning Size of the Sliding Window}

\begin{figure}[htbp]
\begin{center}
  \includegraphics[width=3.5in]{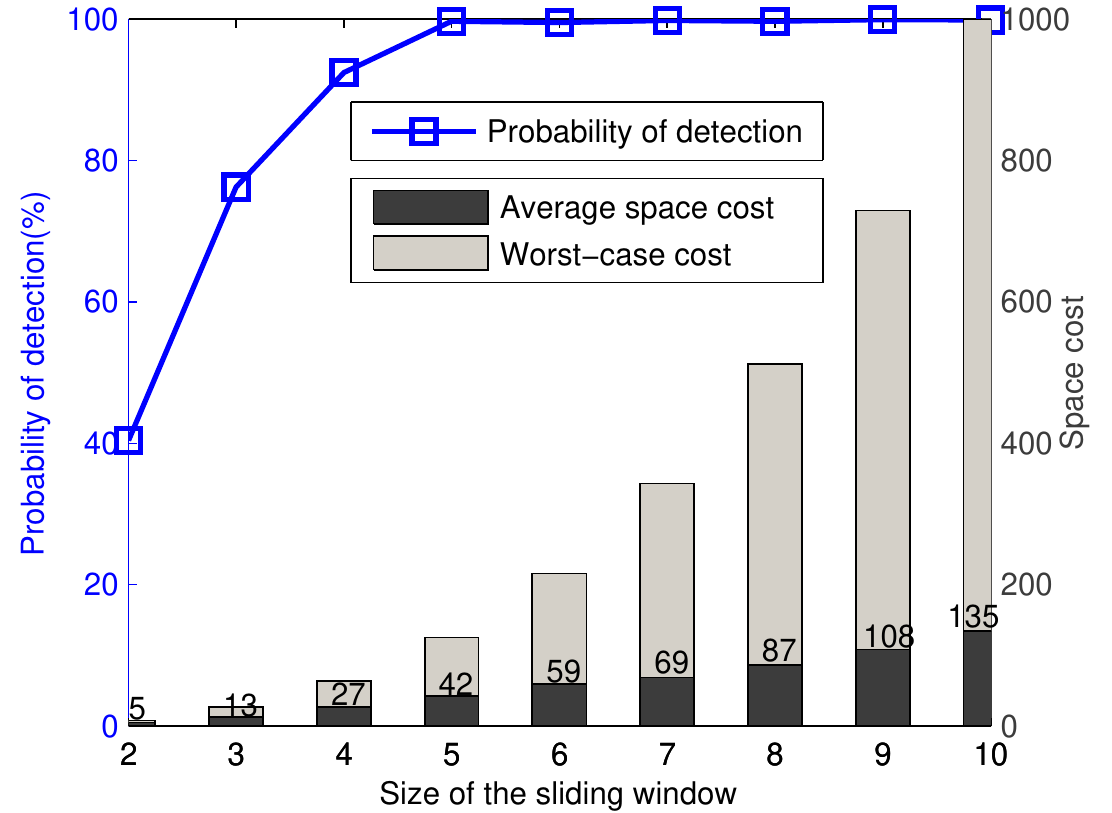}\\
  \centering\parbox[c]{3in}{\caption{Tuning size of the sliding window.}}
  \label{F:window-size}
\end{center}
\end{figure}

In this experiment, we study how the window size affects the performance of {\it Lat-Win}. We fix the message delay to 0.5 s. As shown in Fig. \ref{F:window-size}, the increase of window size $w$ results in monotonic increase in both $Prob_{det}$ and $S_{Lat-Win}$. When we tune $w$ from 2 to 5, $Prob_{det}$ (the upper blue line) increases quickly up to 99\%. The increase is mainly because that as $w$ increases, the window has more information and thus can detect the property more accurately. $S_{Lat-Win}$ (the black bars) increases slowly as $w$ increases, and is much smaller than the worst-case cost. Thus, imposing the sliding window over asynchronous event streams can achieve high accuracy while saving a large amount of space cost. Furthermore, the increase of window size results in a slow increase in $T_{Lat-Win}$, which remains less than 1 ms.

\paragraph{Effects of Tuning the Number of Non-checker Processes}

\begin{figure}[htbp]
\begin{center}
  \includegraphics[width=3.5in]{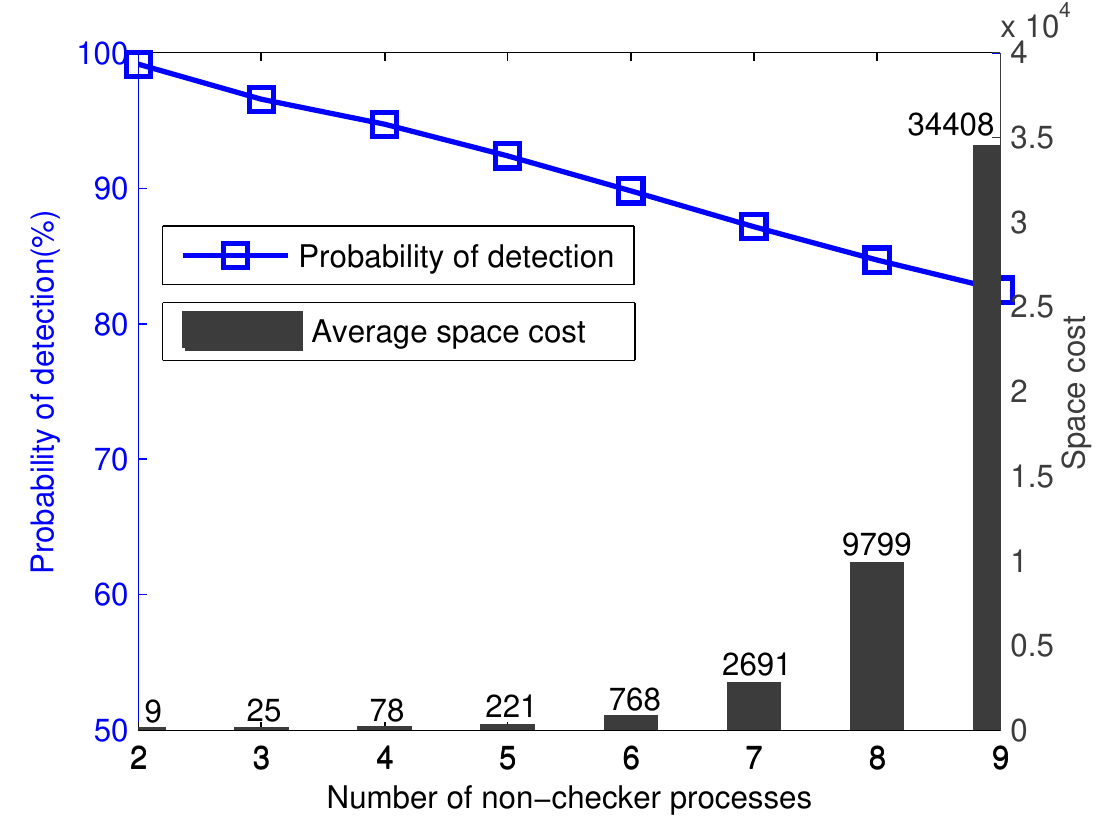}\\
  \centering\parbox[c]{3.9in}{\caption{Tuning the number of non-checker processes.}}
  \label{F:number-of-NPs}
\end{center}
\end{figure}

In this experiment, we study how the number of non-checker processes $n$ affects the performance of {\it Lat-Win}. We fix the average message delay to 0.5 s and the window size $w$ to 4. We tune $n$ from 2 to 9. As shown in Fig. \ref{F:number-of-NPs}, $Prob_{det}$ decreases linearly as $n$ increases. When $n$ increases from 2 to 9, $Prob_{det}$ decreases about 20\%. The decrease is mainly because the asynchrony among non-checker processes accumulates as $n$ increases. $S_{Lat-Win}$ increases exponentially as $n$ increases. However, as $n$ increases, the space cost of {\it Lat-Win} is much less than the worst-case (less than 1\%), and even less than that of the original lattice. $S_{Lat-Win}$ and $T_{Lat-Win}$ are also shown in Table. \ref{T:cost of sliding window}. We find that $S_{Lat-Win}$ is approximately in $O((\theta w)^n)$, which is in accordance with the analysis in Section \ref{sec:complexity analysis}, where $\theta$ is a parameter associated with the asynchrony. In this experiment setting, $\theta$ is around 0.75. $T_{Lat-Win}$ also increases exponentially as $S_{Lat-Win}$ increases.

\begin{table}[htbp]
\caption{Cost of Sliding Window When Tuning the Number of Non-checker Processes}
\label{T:cost of sliding window} \centering
\begin{tabular}{c|c|c|c|c|c|c|c|c}
\hline
Number of NPs ($n$) & 2 & 3 & 4 & 5 & 6 & 7 & 8 & 9 \\
\hline
\hline
$S_{Lat-Win}$ & 9 & 25 & 78 & 221 & 768 & 2691 & 9799 & 34408 \\
\hline
$T_{Lat-Win} (ms)$ & 0.5 & 0.6 & 2.0 & 7.5 & 36 & 184 & 1616 & 14766 \\
\hline
\end{tabular}
\end{table}

\section{Related Work}
\label{sec:related-work}

The lattice of global snapshots is a key notion in modeling behavior of asynchronous systems \cite{Schwarz94, Babaoglu96, Babaoglu95, Cooper91, Hua10, Huang09, Huang10a, Huang11, Mittal07, Sen07}, and is widely used in areas such as distributed program debugging \cite{Garg94, Garg96} and fault-tolerance \cite{Mittal05}. One critical challenge is that the lattice of snapshots evolves to exponential size in the worst-case. Various schemes are used to cope with the lattice explosion problem \cite{Mittal07, Sen07, Dumais02, Jard94, Chen11}. For example, in \cite{Mittal07, Sen07}, the authors proposed the computation slice to efficiently compute all global states which satisfy a regular predicate\cite{Mittal07}. In \cite{Jard94}, the authors discussed that certain (useless) part of the lattice can be removed at runtime to reduce the size of lattice. In this work, we make use of the observation that, in many tracking/monitoring applications, it is often prohibitive and, more importantly, unnecessary to process the entire streams. Thus, we use sliding windows over distributed event streams to reduce the size of the lattice.

Sliding windows are widely used in event stream processing \cite{Babcock02, Braverman11, Datar02, Tirthapura06}. Existing sliding windows are mainly designed over a single stream. However, it is not sufficiently discussed concerning the coordination of multiple (correlated) sliding windows over asynchronous event streams. We argue that the coordination of multiple windows is crucial to explicitly model and handle the asynchrony. In this work, to cope with the asynchrony among multiple event sources, we maintain a sliding window over each event stream. We consider the Cartesian product as an $n$-dimensional sliding window. Then we study the lattice of global snapshots of asynchronous event streams within the window.

\section{Conclusion and Future Work}
\label{sec:conclusion}

In this work, we study the processing of asynchronous event streams within sliding windows. We first characterize the lattice structure of event stream snapshots within the sliding window. Then we propose an online algorithm to maintain {\it Lat-Win} at runtime. The {\it Lat-Win} is implemented and evaluated over MIPA.

In our future work, we will study how to make use of the partial asynchrony among event streams, to further improve the cost-effectiveness of event stream processing. We will also study the approximate/probabilistic detection of specified predicates over asynchronous event streams.

\section*{Acknowledgements}

This work is supported by the National Natural Science Foundation of China (No. 60903024, 61021062) and the National 973 Program of China (2009CB320702).

\bibliographystyle{IEEEtran}
\bibliography{IEEEabrv,TR110211}

\end{document}